\documentclass{article}
\usepackage{lscape}
\usepackage[utf8]{inputenc}
\usepackage{tikz}
\usetikzlibrary{shapes.geometric}
\usepackage{graphicx}
\graphicspath{ {} }
\usepackage{imakeidx}
\usepackage{comment}
\usepackage{amsmath}
\usepackage{framed}
\usepackage{subcaption} 
\usepackage[margin=1in]{geometry}
\usepackage[colorlinks,linkcolor=blue,citecolor=black,urlcolor=black]{hyperref}
\usepackage{natbib}
\title{Testing Piketty’s Hypothesis on the Drivers of Income Inequality: Evidence from Panel VARs with Heterogeneous Dynamics}
\author{Carlos Góes\footnote{University of California - San Diego. I wholeheartedly thank Branko Milanovic, Peter Pedroni, Daron Acemoglu, Marcelo Estevão, Alfredo Cuevas, Roberto Perrelli, Henrique Barbosa, Franz Loyola, Troy Matheson, Iza Karpowicz, Michele Andreolli, Alexandra Martinez, and participants in IMF and University of Brasilia seminars for their helpful comments. All errors and omissions are solely mine. This paper is fully reproducible: Data and econometric programs are available at \url{https://github.com/omercadopopular/cgoes}. The author can be contacted by email at  \href{mailto:cgoes@ucsd.edu}{cgoes@ucsd.edu}.}}

\begin{document}

\maketitle

\begin{abstract}
\noindent Thomas Piketty's \textit{Capital in the Twenty-First Century} puts forth a logically consistent explanation for changes in income and wealth inequality patterns. However, while rich in data, the book provides no formal empirical testing for its theorized causal chain. This paper tests the hypothesis that the $r-g$ gap drives income inequality and the increasing capital share of national income. Using panel VAR models with data from 18 advanced economies over 30 years, I find no empirical support for Piketty's predictions. The results suggest that dynamics such as savings-rate adjustments and diminishing returns to capital play critical roles in offsetting the hypothesized effects. These findings challenge the theoretical underpinnings of the growth in inequality and call for alternative explanations.
\end{abstract}

\newpage

\section{Introduction}

\textit{Capital in the Twenty-First Century}, Thomas Piketty's magnum opus, has been widely praised for aggregating and presenting in a timely and accessible fashion the results of more than a decade of research  by him and his coauthors. His work is likely to remain influential, not only for its databases, which systematize (traditionally scarce) information about income inequality, but also because it is inspiring many economists to use the estimation techniques he and his coauthors popularized to further our knowledge of these phenomena in different countries.

Piketty's theoretical explanations for changes in inequality patterns are logically consistent. Using a standard neoclassical model, he argues that all other things constant, whenever the difference between the return on capital ($r$) and the output growth rate ($g$) increases, the share of capital in the national income must increase. Furthermore, since capital income tends to be more unequally distributed than labor income, an increase in the capital share would likely lead to increased overall income (and, over time, wealth) inequality. Both of these are plausible relationships. However, while rich in data, \textit{Capital} provides no formal empirical testing for these conjectures. 

The main contribution of this paper is to provide an empirical test of Piketty's hypotheses. My findings indicate that changes in $r-g$ do not lead to increased  capital shares or inequality  in most advanced economies. This evidence suggests that other mechanisms, such as labor market shifts and institutional factors, play a more significant role. These results challenge the policy prescriptions derived from Piketty's framework and invite a reexamination of its theoretical foundations.

I test whether inequality and the capital share of  national income increase as the $r-g$ gap grows using structural panel vector autoregressive models (VARs). I estimate panel VARs as described in \cite{pedroni}, which allows me to control for country fixed effects and capture fully heterogeneous dynamics across countries. This means that rather than just capturing average effects, I estimate a distribution of impulse response functions, permitting  much more robust inference than those that rely on average estimates and assume slopes are homogeneous.

I find no empirical evidence that the dynamics are as Piketty suggests. In fact, for at least 75\% of the countries examined, inequality responds negatively to $r-g$ shocks, which is in line with previous single-equation estimates by \citet{acemoglu}. The results also suggest that changes in the savings rate, which Piketty takes to be relatively stable, are likely to offset most of the impact of $r-g$ shocks on the capital share of the national income. Thus, they provide empirical support for  the theoretical model developed by \citet{krusell}, who say Piketty relies on a flawed theory of savings. The conclusions are robust to alternative estimates of $r-g$ and to the exclusion or inclusion of tax rates in the calculation of the real return on capital.

Knowing whether Piketty's hypothesis is correct is crucial because the policy solutions designed to counter increasing income inequality  in advanced economies will need to tackle its underlying causes. The results presented here show that observed increases in income inequality in advanced economies are largely uncorrelated with changes in $r-g$, which suggests that one needs to look for the causes of inequality (and potential solutions) elsewhere.

This paper outlines Piketty's theoretical model and develops a strategy to test it. First, I present a simplified description of the basic theoretical relationships Piketty proposes. Then, I describe the data and how I constructed some of the variables of interest and explain the empirical methodology applied. Afterward, I present the results (paying special attention to how to interpret heterogeneous dynamics), perform some robustness tests, and  relate the results to the rest of the literature on income inequality, pointing to potential causes of increasing inequality that are not related to $r-g$. Finally, I conclude by summarizing the most important points of the paper and what they mean for future inequality dynamics.

\section{Piketty's Model: What to Test}\label{sec: model}

Building on a standard growth model, \citet*{piketty1} argues that the patterns of wealth and income concentration are defined by the difference in the real return on capital ($r$) and the growth rate ($g$). Here, I present a very stylized derivation of Piketty's model and its implications so as to develop a testing strategy. 

Envision a closed economy where national income ($Y$) is a function of capital ($K$) and labor ($L$), $Y_t = K_t^{\alpha} (A_t L_t) ^{1 - \alpha}$, with $A_t$ being a time-varying labor-augmenting technology, and the final good is produced in competitive markets. Under these assumptions, an equilibrium implies that factor prices equal their marginal products: $r_t = \frac{\partial Y_t}{\partial K_t} = \alpha Y_t / K_t$, $w_t = \frac{\partial Y_t}{\partial L_t} = (1-\alpha) Y_t / L_t$. By inverting the first-order condition with respect to capital, Piketty derives what he calls the ``first fundamental law of capitalism'':

\begin{eqnarray} \label{eq:alpha}
    \alpha = \frac{r K}{Y}
\end{eqnarray}

Assume further that investment is a constant share $s$ of national income in any period $t$,  as in the standard \cite{solow1956contribution} model. Under that assumption, the law of motion of capital is $K_{t+1} = (1 - \delta) K_{t} + sY_{t}$. Suppose that $A_t$ grows exogenously at rate $(1+g)$ and population is constant. Then, one can show that, over the balanced growth path (BGP), $\frac{Y_{t+1}}{Y_{t}} = \frac{K_{t+1}}{K_{t}} = \frac{A_{t+1}}{A_{t}} = (1+g)$. Using the law of motion for capital, then, along the BGP,\footnote{Piketty defines all of his variables in \textit{net} terms, such that one should deduct depreciation from income, capital, and the savings rate. But since both expressions are equivalent in their steady states \citep{krusell}, and since most standard textbooks use variables in gross terms, it is more expedient to explicitly account for depreciation.}

\begin{equation} \label{eq:kdynamics}
     (1+g) K_t = (1-\delta) K_t + sY_t \iff \frac{K_t}{Y_t} = \frac{s}{g + \delta}.
\end{equation}

Substituting (\ref{eq:kdynamics}) into (\ref{eq:alpha}) yields what Piketty calls the ``second fundamental law of capitalism''\textemdash an inverse relationship between the share of capital in  national income and economic growth:

\begin{eqnarray} \label{eq:pik}
    \alpha = \frac{r s}{g + \delta}
\end{eqnarray}

\noindent Here, $r$ denotes the real return on capital along the BGP.\footnote{For instance, in a standard model with CRRA preferences, one can use the Euler equation and find that $r = \frac{1 }{\beta} - (1-\delta)$.}

Taking the (net) savings rate as somewhat constant, \citet{piketty2} argue that the capital share, income inequality, and wealth inequality are rising functions of $r-g$. If Piketty is correct, then one should expect a positive shock to $r$ or a negative shock to $g$ to increase the capital share of national income. Piketty goes on to argue that as the returns on capital are more unequally distributed than labor income, a higher share of capital in national income would lead to higher income and wealth inequality \cite[ch.~7]{piketty1}. 

These relationships are empirical propositions and hence empirically testable. The testable hypotheses are that positive changes in $r-g$ lead to positive changes in inequality and the capital share. Piketty stresses that these dynamics are long run and asymptotic, hinting at potential difficulties in capturing them. Still, if the model is correct, deviations around the long-run average should provide some evidence of the underlying relationships. 

\section{Data and Stylized Facts}

To analyze the relationship between inequality, the capital share, and the  differential between the return on capital and GDP growth, I gather data from multiple sources. Data availability varies by country, and the maximum range used in this paper goes from 1980 through 2012. For this longer, data on the capital share and share of the top 1\% are more reliable for advanced economies relative to most developing countries. For this reason, the final sample is an unbalanced long panel with annual observations for 18 advanced economies from different continents (see Appendix~\ref{appendix:appendix1} for details).

Inequality is proxied by the share of national income held by the top 1\%, as reported by the World Top Incomes Database. I choose this variable  since this is Piketty's measure of choice (rather than  alternative metrics, such as the Gini index) indicating rising inequality in advanced economies in recent decades. The annual capital share of national income comes from the Penn World Tables. 

To derive the second variable of interest\textemdash namely, the real return on capital net of real GDP growth\textemdash I take yearly averages of nominal long-term sovereign-bond yields, calculate the post-tax nominal rates by deducting corporate income taxes, and subtract from them annual percent changes in GDP deflators and real GDP growth. The $r$ in $r - g$ is, then,

\begin{eqnarray} \label{eq:rr}
    r_{i,t} = [ (1 - \tau_{i,t}) i_{i,t} - d_{i,t} ],
\end{eqnarray}

\noindent where $r_{i,t}$ is the real return on capital, $\tau_{i,t}$ is the corporate income tax rate, $i_{i,t}$ is the nominal long-term sovereign-bond yield, and $d_{i,t}$ is the annual percent change in the GDP deflator for country $i$ in  period $t$. GDP deflators and real growth rates are found in the International Monetary Fund's World Economic Outlook database. Tax rates come mostly from the OECD's Tax Database\textemdash with the exception of the time series for Singapore, which was constructed independently. 

The reason for choosing sovereign-bond yields as a proxy for returns on capital is not self-evident. In reality, the aggregate return on capital is a weighted average of returns across a plethora of investments. However, returns on government bonds are a good proxy for the purposes of this empirical exercise for two reasons. First, Piketty's narrative centers on worries about a society facing the unmeritocratic rule of rentiers who profit from high returns on capital, which he illustrates by discussing how 18th-century élites profited from high returns on government bonds \citep[p.~132]{piketty1}. Second, even if the level of return is different for different portfolios, the correlation between sovereign bonds and corporate bonds is historically very high. In fact, between 1996 and 2015, the correlation between prime AAA corporate bonds and US Treasuries was nearly perfect ($r = 0.93$); even when considering ``junk'' (BB+ graded) corporate bonds, the co-movement between sovereigns and corporates was relatively tight ($r = 0.49$).

\citet*{piketty6} criticized  an earlier version of this paper for the use of interest rates as a proxy for returns on capital. His argument is a sensible one: In modern times, portfolios are very diversified, and the returns on capital for the very wealthy are often higher than returns on government bonds. He referred to his own estimates of the real wealth growth rate of the top 50,000 individuals, based on \textit{Forbes} magazine's list of billionaires, to argue that it outpaced overall per capita real wealth growth in the 1987--2013 period. But there is high uncertainty regarding the quality of the \textit{Forbes} estimates. For instance, \citet{raub2010comparison} find that wealth reported in tax data only corresponds to about half of \textit{Forbes}'s list. With different assumptions, either number could be closer to the true stock of wealth. 

More importantly, this argument does not address whether  interest rates are a good proxy for returns on capital. Level differences notwithstanding, insofar as returns on capital can be reasonably expressed as a linear function of returns on government bonds, the latter are a good proxy for the former. 

Data on heterogeneous returns on capital are scarce. Some years ago, the editors of  Credit Suisse's \textit{Global Wealth Databook}, who have been publishing estimates of wealth concentration across countries for several years, highlighted that the  ``study of global household wealth is still in its infancy'' \citep{creditsuisse}. More recently, the \textit{World Inequality Report} recalled that ``global wealth data remain opaque,'' and many of the estimates they provide come from extrapolations from income inequality data or limited population coverage.

These caveats notwithstanding, the Federal Reserve  recently published its Distribution Financial Accounts. They provide estimates of asset stocks for different ownership brackets going back to 1989. The database uses underlying microdata from the Survey of Consumer Finances. Regressing the four-quarter net return on assets\footnote{If $A_{q,t}^{1\%}$ is the asset stock of the top 1\% in quarter $q$ of year $t$, the four-quarter net return on assets is $R_{q,t}^{1\%} = \frac{\sum_{p=0}^3 A_{q-p,t}^{1\%} }{ \sum_{p=0}^3 A_{q-p,t-1}^{1\%} } -1$. } of the top 1\% on three-month Treasuries yields the following relationship: $\hat{R}_{q,t}^{1\%} = \underset{(0.010)}{1.81} + \underset{(0.000)}{1.60} R^{3mo}_{q,t}$. The relationship with 10-year bonds is $\hat{R}_{q,t}^{1\%} = \underset{(0.93)}{2.23} + \underset{(0.003)}{1.30} R^{10yr}_{q,t}$.

The specific criticism Piketty made regarding the use of sovereign-bond yields thus appears to be largely unfounded. However,  recognizing the challenge to find good proxies for the return on capital, I present results with two different proxies for $r$ in Section~6: short-term interest rates, and implied returns from the national accounts. I also present results ignoring tax rates. 

\begin{figure}[htp]
\begin{center}
    \includegraphics[scale=0.6]{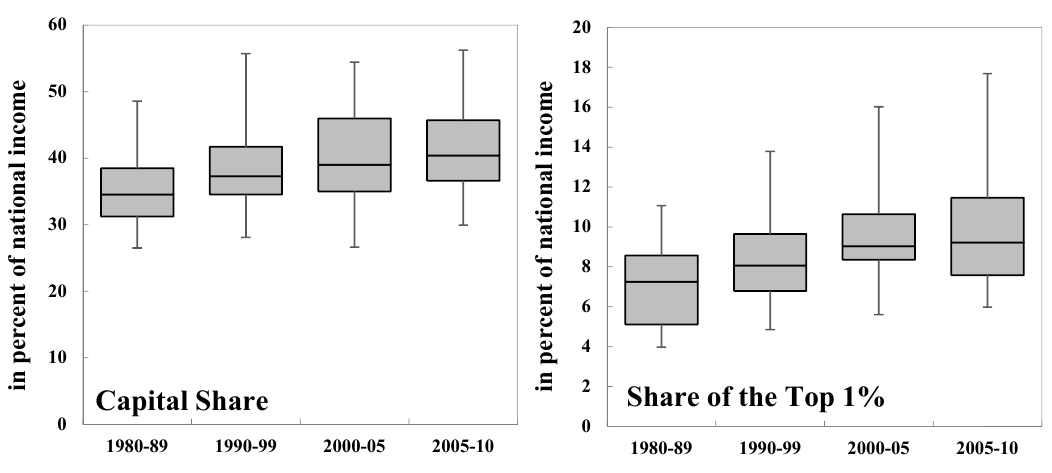}
    \caption[Distribution of capital share and share of the top 1\% over time]{\textbf{Distribution of capital share and share of the top 1\% over time.} Y-axis in percent, x-axis represents period averages. The sample refers to an unbalanced panel of 18 advanced economies ranging from 1981 to 2010. Boxplots show interquartile ranges and medians. Whiskers show minimums and maximums.} \label{fig:dist}
\end{center}

\end{figure}

Distributions of the capital share and share of the top 1\% in the sample show increasing trends, although with some signs of moderation in the late 2000s due to the Global Financial Crisis. The upward trend is observed not only in the medians but also in the evolution of interquartile ranges, as shown in Figure~\ref{fig:dist}.

\begin{figure}[htp]
\begin{center}
    \includegraphics[scale=0.6]{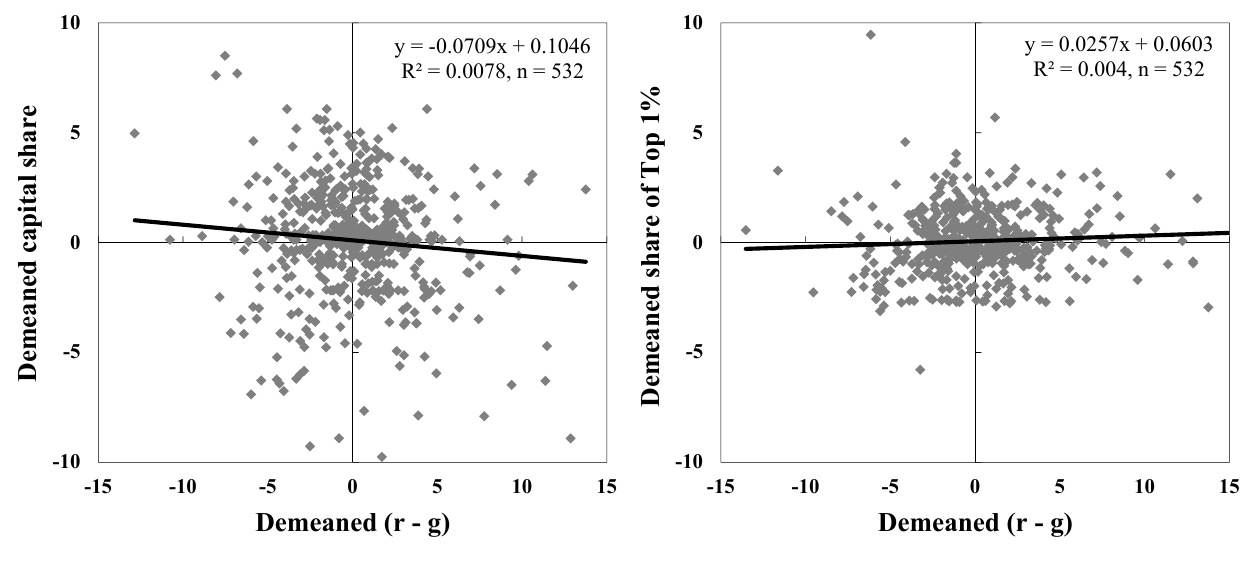}
    \caption[Contemporaneous correlations between $r-g$ spread and capital share or share of the top 1\%]{\textbf{Contemporaneous correlations between $r-g$ spread and capital share or share of the top 1\%, respectively.} The sample refers to an unbalanced panel of 18 advanced economies ranging from 1981 to 2012. Variables are de-meaned to account for time-invariant country-specific characteristics.} \label{fig:corr}
\end{center}
\end{figure}

Stylized facts can provide some preliminary insights regarding whether  the data support Piketty's assertions. Figure~\ref{fig:corr} plots the contemporaneous correlations between $r - g$ spreads and the capital share, and between the spreads and the share of the top 1\%. Such  correlations show no evidence of the relationship Piketty poses. Rather, the variables seem largely orthogonal. However, as mentioned before, Piketty's theory is largely about the long run.

If the absence of a positive relationship persists with more robust methods, this could raise questions about Piketty's assertion that  higher $r - g$ will lead to higher inequality over the coming century. In the next section, I describe an empirical method to test Piketty's central hypothesis.

\section{Methodology}

Following \cite{pedroni}, I estimate a structural panel VAR that accommodates country fixed effects and allows dynamics to be fully heterogeneous among panel members. Finding heterogeneous dynamics means estimating country-specific coefficients and impulse response functions, rather than estimating average slopes.

A full description of the methodology is in Appendix~\ref{appendix: methodology}. To save space, I describe it only briefly here. The Pedroni method consists of (i) estimating individual structural VARs for each member of a panel, accounting for individual fixed effects, extracting ``composite shocks'';   (ii) estimating an auxiliary VAR for the cross-sectional averages of the data, extracting ``common shocks''; (iii) estimating the correlations between the estimated common and individual composite shocks to consistently decompose dynamics between those driven by common versus idiosyncratic shocks; and (iv) describing moments of the cross-sectional distribution of composite, common, and idiosyncratic impulse responses (one can plot, for instance, the average, median, and interquartile range of responses).
    
For each member $i \in \{1, ..., N\}$ of an unbalanced panel, there is a vector $y_{i,t}$ of $M$ endogenous variables with country-specific time dimension $t \in \{1, ..., T_i\}$. To control for individual fixed effects, all data are de-meaned as $y^*_{i,t} = y_{i,t} - \bar{y}_{i} $, where $ \bar{y}_{i} \equiv T_i^{-1} \sum_{t=1}^{T_i} y_{i,t}    \forall   i$.
    
I run a structural VAR for every $i$, with  $y^*_{i,t}$ as the endogenous variable. All variables are endogenous in the sense that lags of every variable are assumed to causally affect all other variables in the system. However, to identify the \textit{contemporaneous relationships} between the $M$ endogenous variables, one needs to impose $M(M-1)/2$ restrictions on the structure of contemporaneous shocks. Here, I use the standard methods for performing a Cholesky decomposition in the variance-covariance matrix, assuming that the variable ordered first ($m=1$) affects all other variables but is unaffected by other variables contemporaneously. Imposing this structure, any endogenous variable $m$ affects all variables $k>m$ contemporaneously. \cite{lutkepohl} describes how to use the matrix of contemporaneous relationships to recover structural impulse response functions. 
    
The model also decomposes responses according to the contribution of \textit{idiosyncratic} and \textit{common} shocks. Common shocks capture global events that induce macroeconomic variables to vary in a correlated fashion across countries, such as a global pandemic or  recession. Idiosyncratic shocks capture country-specific  unpredictable events unrelated to global or shared trends. Intuitively, if the common-component contribution is large, the correlation between average shocks and country-specific shocks is large. Conversely, if this correlation is low, then dynamics are dominated by the idiosyncratic component.
    
To perform this decomposition, I estimate a structural VAR using as the endogenous vector the cross-sectional average for each period ($\bar{y}^*_{t} \equiv N^{-1} \sum_{i=1}^{N} y^*_{i,t}$). After recovering individual composite structural residuals for each country $i$ (from country-specific VARs) and common structural residuals (from the VAR estimated with cross-sectional averages), I  calculate the correlation between common and composite residuals for each country. These are  loadings that allow me to decompose the responses in each country between those driven by common shocks and those driven by idiosyncratic shocks. 

\section{Empirical Models and Results}

I run three models with the methodology described above:

\begin{itemize}
    \item In \textbf{Model 1}, $y_{i,t} \equiv [p_{i,t}, z_{i,t}]'$, where $z_{i,t}$ is the national income share of the top 1\% for country $i$ in period $t$ and $p_{i,t} \equiv (r_{i,t} - g_{i,t})$ is the post-tax return on capital ($r$) net of real GDP growth ($g$).
    \item In \textbf{Model 2}, $y_{i,t} \equiv [p_{i,t}, k_{i,t}]'$, where $k_{i,t}$ is the share of capital in national income.
    \item In \textbf{Model 3}, $y_{i,t} \equiv [p_{i,t}, s_{i,t}, k_{i,t}]'$, incorporating the savings rate $s_{i,t}$.
\end{itemize}

Since the data are de-meaned to control for country fixed effects, if one takes the rate of depreciation as time invariant, even if different for each country, it is not necessary to explicitly account for it in the empirical model. A time invariant depreciation rate is accounted for once the data are de-meaned as $p^*_{i,t} \equiv (r_{i,t} - g_{i,t})^* = (r_{i,t} - g_{i,t} - \delta_i)^*$.

In identifying the Cholesky decomposition, I order $r-g$ first, imposing lower triangular restrictions in the contemporaneous coefficients such that (i) $r-g$ contemporaneously affects the share of the top 1\% and the capital share and (ii) $r-g$ responds to shocks to the share of the top 1\% and the capital share only with a lag. In Section~6, I present results with the inverse restrictions. 

\begin{figure}[htp!]
\begin{center}
        \includegraphics[width=\textwidth]{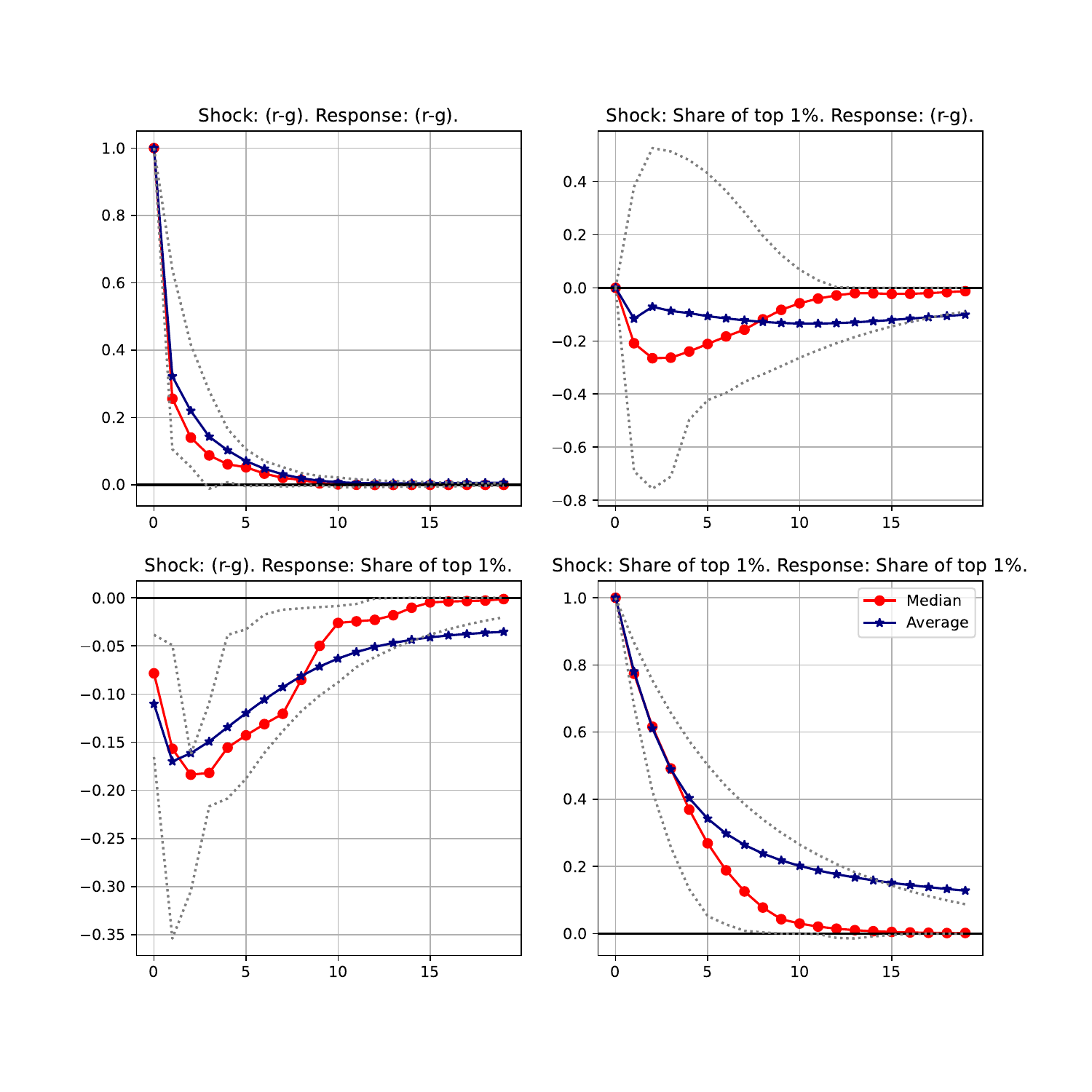}
\end{center}
\caption[Results of Model 1: Heterogeneous composite impulse responses across sample]{\textbf{Model 1: Heterogeneous composite impulse responses across sample.} The median and averages are shown as solid lines, interquartile ranges as dotted lines. All are calculated from the distribution of impulse response functions of the 18 cross-sections.}\label{fig:top1-dist}
\end{figure}

Pedroni's methodology allows for fully heterogeneous dynamics. This means that the results of the panel VAR are more than \textit{average} parameters and \textit{average} impulse response functions\textemdash unlike in traditional panels, which impose homogeneous parameters. Rather, I have information about several moments of the distributions of impulse responses for each response horizon. With those data, I can then plot averages, medians, and interquartile ranges of responses for a given horizon, as shown in Figure~\ref{fig:top1-dist}.

This is a much more informative way of reading results than in traditional panel VAR analyses. For instance, had I calculated average impulse responses from parameters estimated with traditional dynamic panels (for example,  difference or system GMM equations), I would have no way of knowing how many countries in the sample have dynamics that are similar to the average dynamics\textemdash as the underlying assumption is that parameters are equal for all countries. Knowing exactly how many countries in the sample present certain dynamics allows for much more robust inference than simply relying on average estimates. In addition, as shown by \citet{pesaran}, if individual dynamics are heterogeneous, aggregating or pooling slopes can lead to inconsistent estimates, making individual regressions for each group member preferable.

Despite country-specific heterogeneity, the estimated VARs are stable, which we know because  the inverse roots of the characteristic polynomial related to each country-specific matrix of VAR coefficients are within the unit circle. Several of the variables are shares and have bounded support, which rules out the possibility of a unit root in the traditional sense.\footnote{\citet{granger2010some} extends the idea of nonstationarity to bounded variables. He calls a stochastic process that behaves as a $I(1)$ process over limited samples but are stationary over the long run a \textit{limited integration } (or $LI(1)$) process. For many of the variables I use in the estimation, there are upper and lower bounds, such as capital share or the savings rate, both bounded within $[0,1]$. Where variables are bounded, by assumption, those variables cannot be nonstationary in the traditional sense but could behave like a nonstationary variable away from the bounds.} Others, such as $r-g$, must have some finite mean for an economic equilibrium to exist. 

This means that one-off shocks should be interpreted as temporary and, following any shock, variables are expected to converge on their means or deterministic trends in the long run. In line with the outlined theory, one should interpret these as deviations around the long-run BGP. Permanent shocks are interpreted as deviations in the long-run relationships  along the BGP or changes in the long-run equilibrium growth rate. Through the lens of the stylized model presented in Section~\ref{sec: model}, a permanent shock in the savings rate moving it to $s' > s$ would imply a permanent increase in the capital-to-output ratio. A permanent negative shock to $g$ (which, in turn, is a positive shock to $r-g$) would imply a permanent increase in the capital-to-output ratio and a change in the deterministic trend. In that sense, a permanent shock would mean a transition across BGPs and could also be recovered by accumulating the marginal impulse response functions presented below. I present accumulated impulse response functions in Appendix~\ref{appendix: ac_irfs}. All of the qualitative results described in this section remain unchanged whether referring to marginal or accumulated responses.
 
\begin{figure}[htp!]
\begin{center}
    \includegraphics[width=\textwidth]{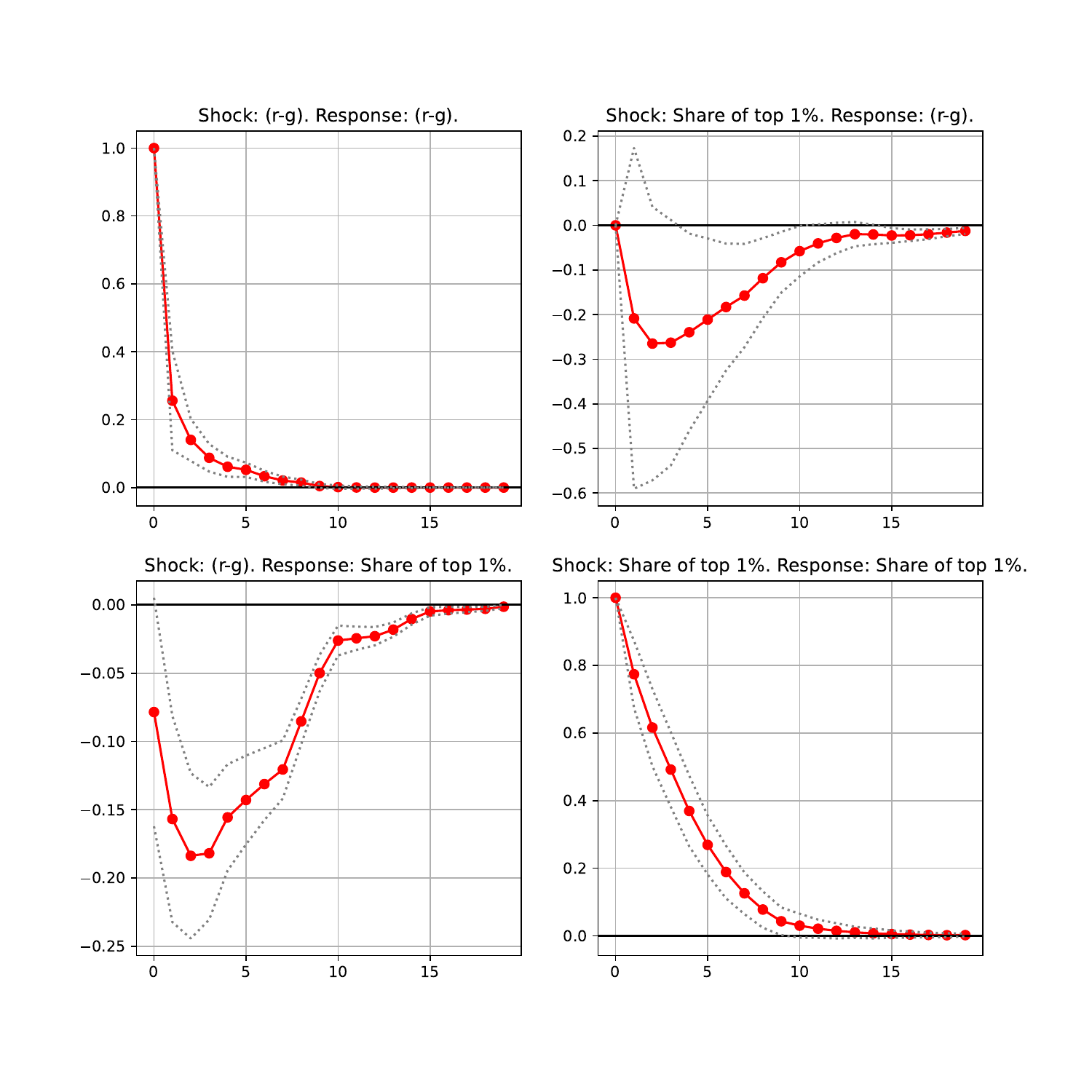}
\end{center}
\caption[Results of Model 1: Median composite responses and confidence intervals]{\textbf{Model 1: Median composite responses and confidence intervals.} Median response across a heterogeneous distribution of impulse response functions across 18 countries. Confidence intervals calculated from a resampling simulation with 1,000 repetitions. Since the distribution might change for each repetition in the simulation exercise, the confidence intervals do not represent the uncertainty around estimates for any particular country, but rather for the median estimate of the whole panel.}\label{fig:top1-ci}
\end{figure}

\begin{figure}[htp!] 
\begin{center}
    \includegraphics[width=\textwidth]{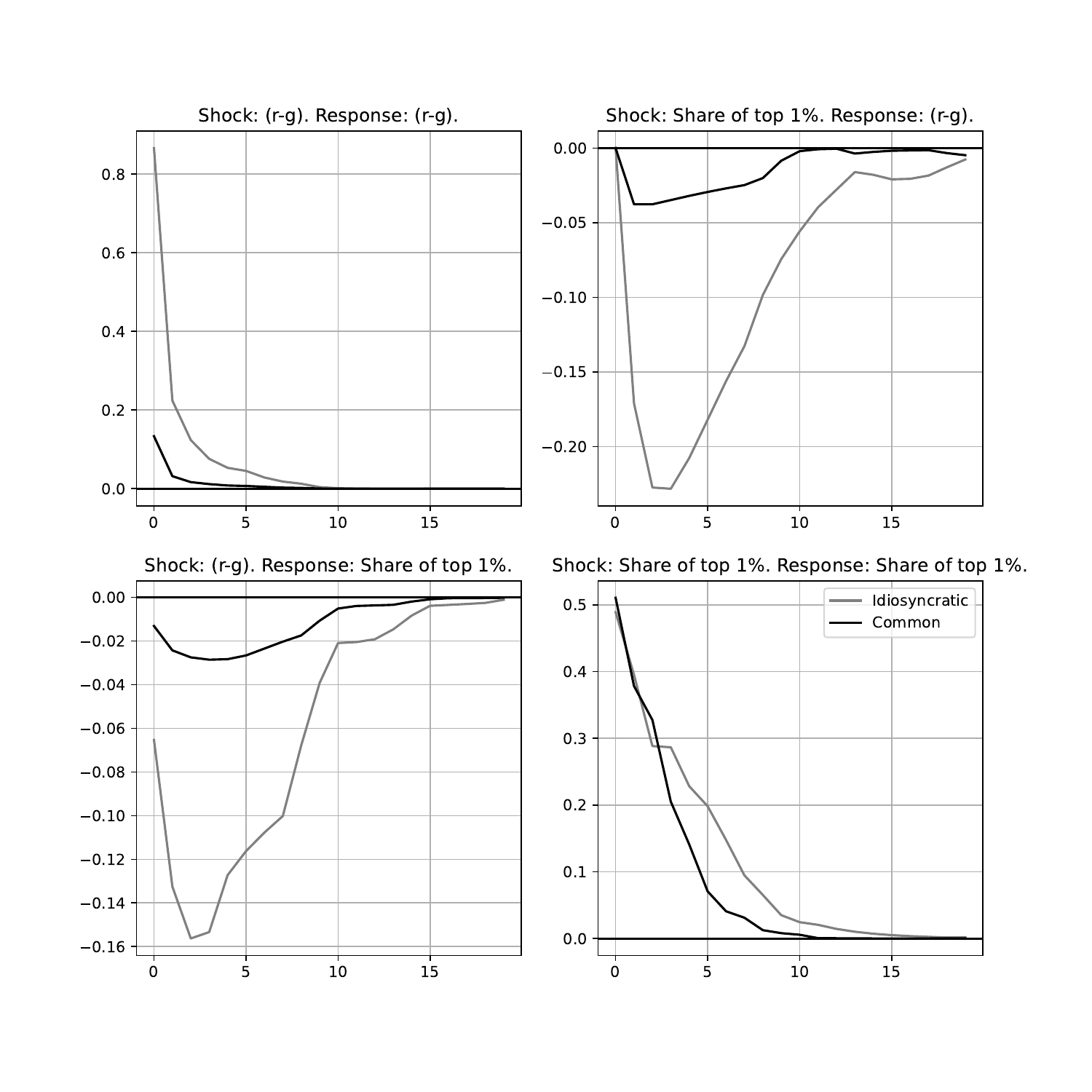}
\end{center}
\caption[Results of Model 1: Decomposition of median composite responses]{\textbf{Model 1: Decomposition of median composite responses.} Median responses can be decomposed into country-specific responses to common shocks and responses to idiosyncratic shocks through the use of loading factors that denote the relative importance of common shocks for each country.}\label{fig:top1-decomp}
\end{figure}

The results of Model~1 show that for at least 75\% of the countries, a 1~pp positive innovation to $r - g$ leads to an expected decrease in the share of the top 1\% (see Figure~\ref{fig:top1-dist}). While the median response is not statistically significant contemporaneously, it is negative and statistically significant in the medium to long run (see Figure~\ref{fig:top1-ci}). The responses to common and idiosyncratic shocks have the same (negative) sign and contribute to the effect with about the same response. This decomposition can help one understand how much systemic forces (such as globalization) are  exacerbating inequality as returns on capital increase (see Figure~\ref{fig:top1-decomp}).

\newpage

\newpage

\newpage

One of the advantages of estimating heterogeneous slopes is that one can explore the variability in parameters across the cross-sectional dimension of the panel. One can do that by correlating the response magnitudes of the different countries in the sample with potential covariates. I plot below some bivariate relationships that can provide some intuition on the causes and effects of the  heterogeneity in responses.

There is a positive correlation between estimates of the intergenerational elasticity of income\textemdash which measures the share of a person's income explained by the parent's income\textemdash and the size of the contemporaneous response of inequality to $r-g$ shocks (Pearson's $\rho = 0.40$). This hints that a higher transmission between $r-g$ and the share of the top 1\% leads to a higher observed level of wage and wealth persistence. 

\begin{figure}[htp]
\begin{center}
    \includegraphics[width=\textwidth]{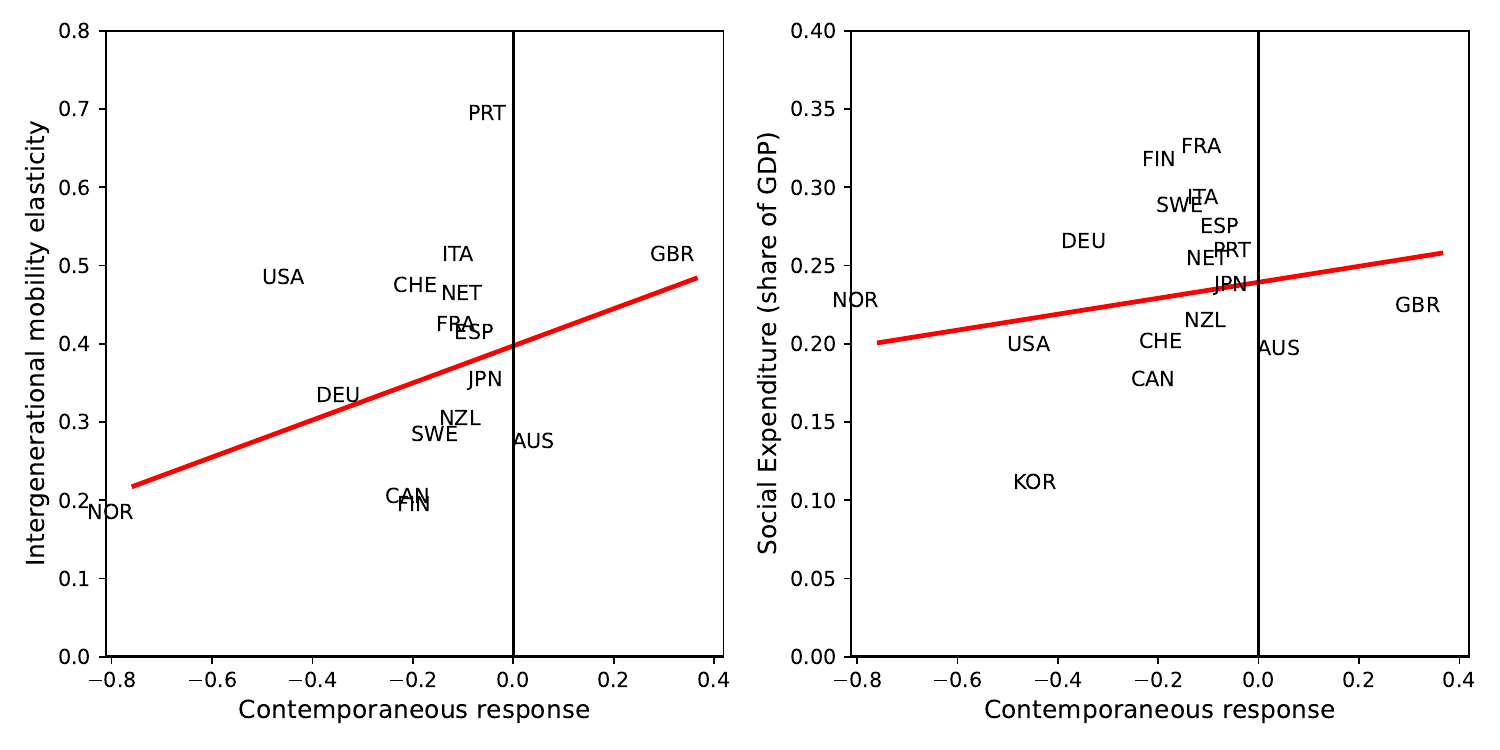}
    \caption[Correlation between heterogeneous responses and intergenerational elasticity of income and social expenditure, respectively]{\textbf{Correlation between heterogeneous responses and intergenerational elasticity of income and social expenditure, respectively.} Data on intergenerational elasticity of income from Corak \citep*{corak} and Causa and Johansson \citep*{causa}. Data on social spending from the OECD's social-expenditure database.} \label{fig:cross-sections}
\end{center}
\end{figure}

Similarly, contemporaneous responses of inequality to $r-g$ shocks are positively correlated with social expenditure as a share of GDP for countries in the sample. This suggests that societies that are more vulnerable to larger shocks to inequality tend to be more willing to resort to government to provide social services (Pearson's $\rho = 0.23$). This is a Rawlsian interpretation of how to hedge one's exposure to inequality under a  veil of ignorance  about the future: Insofar as social services are usually targeted at the lower end of the income distribution, they can be seen as a kind of insurance against higher market-income inequality. Such correlations are shown as simple bivariate scatterplots in Figure~\ref{fig:cross-sections}.

However, these results need to be read with caution since the regressions' slopes are not statistically significant at standard critical levels. This could be either a byproduct of the small cross-sectional dimension of the sample or an indication that in a broader sample these results would not hold. Exploring further the causes and consequences of the heterogeneity in the relationship between $r-g$ dynamics and inequality is a potential avenue for future research. 

I now turn to Model~2, which checks the relationship between $r-g$ and the capital share. This specification can be thought of a direct test of Piketty's model  since his extrapolations to inequality stem from  growth in the capital share. 

\begin{figure}[htp!]
\begin{center}
    \includegraphics[width=\textwidth]{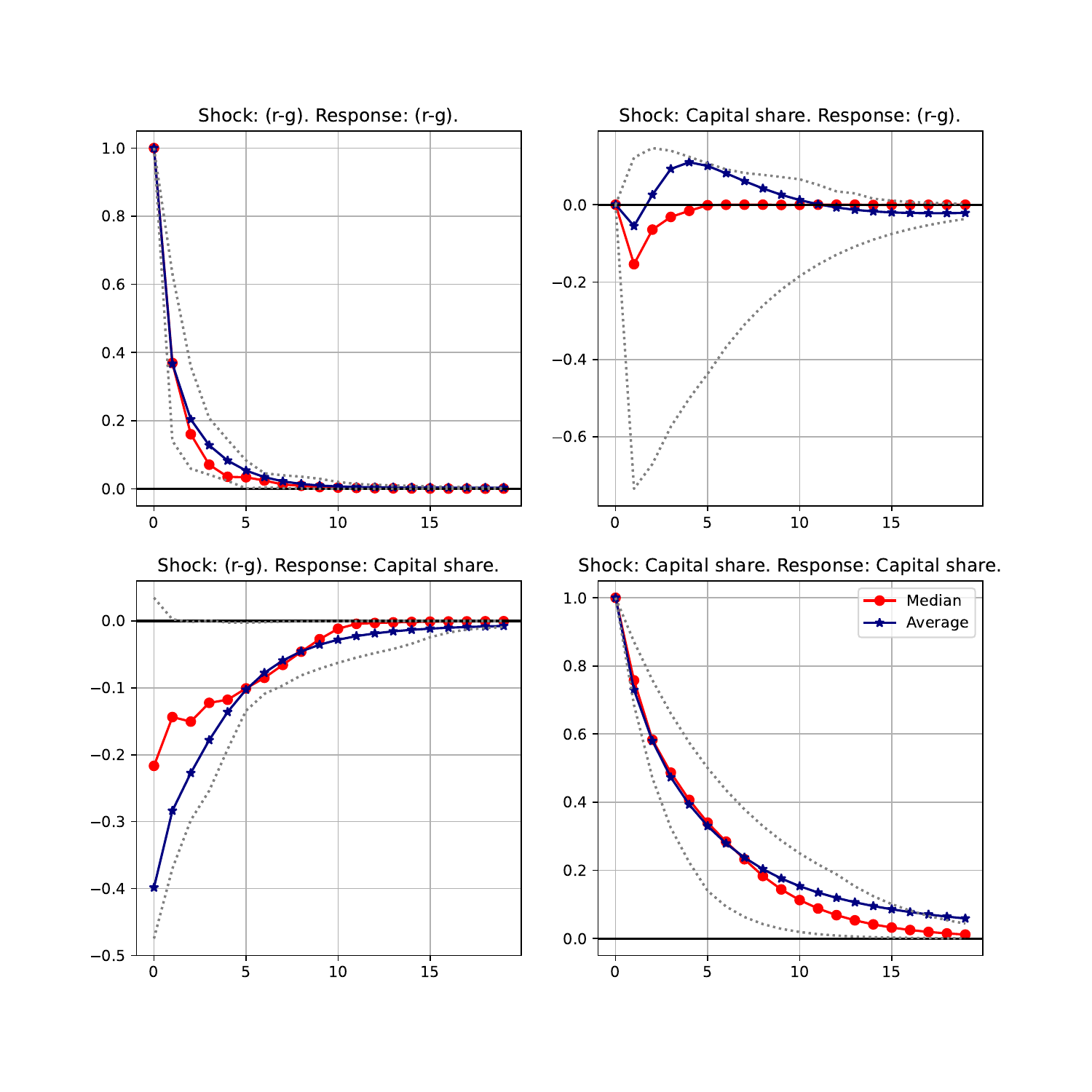}
\end{center}
\caption[Results of Model 2: Heterogeneous composite impulse responses across sample]{\textbf{Model 2: Heterogeneous composite impulse responses across sample.} The median and averages are shown as solid lines, interquartile ranges as dotted lines. All are calculated from the distribution of impulse response functions of the 18 cross-sections.}\label{fig:k-dist}
\end{figure}

\begin{figure}[htp!]
\begin{center}
    \includegraphics[width=\textwidth]{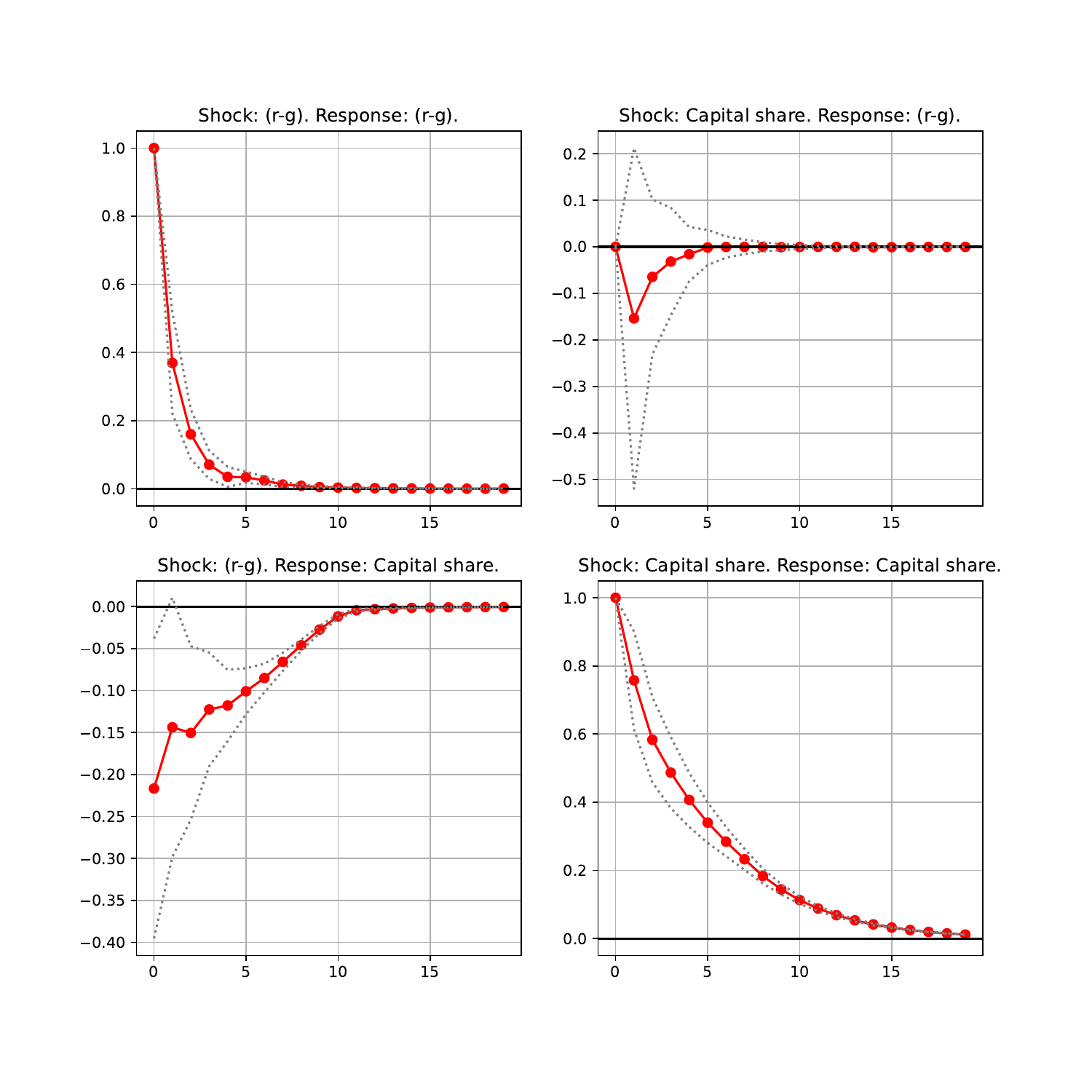}
\end{center}
\caption[Results of Model 2: Median composite responses and confidence intervals]{\textbf{Model 2: Median composite responses and confidence intervals.} Median response across a heterogeneous distribution of impulse response functions across 18 countries. Confidence intervals calculated from a resampling simulation with 1,000 repetitions. Since the distribution might change for each repetition in the simulation exercise, the confidence intervals do not represent the uncertainty around estimates for any particular country, but rather for the median estimate of the whole panel.}\label{fig:kirf-ci}
\end{figure}

The results are less clear-cut than with Model~1. However, in the first two years, at least 75\% of the countries' capital shares respond (mildly) negatively to positive shocks to $r-g$. Median responses are substantially negative, statistically significant on impact, and approach zero asymptotically (see Figures~\ref{fig:k-dist} and \ref{fig:kirf-ci}).

\begin{figure}[htp!]
\begin{center}
    \includegraphics[width=\textwidth]{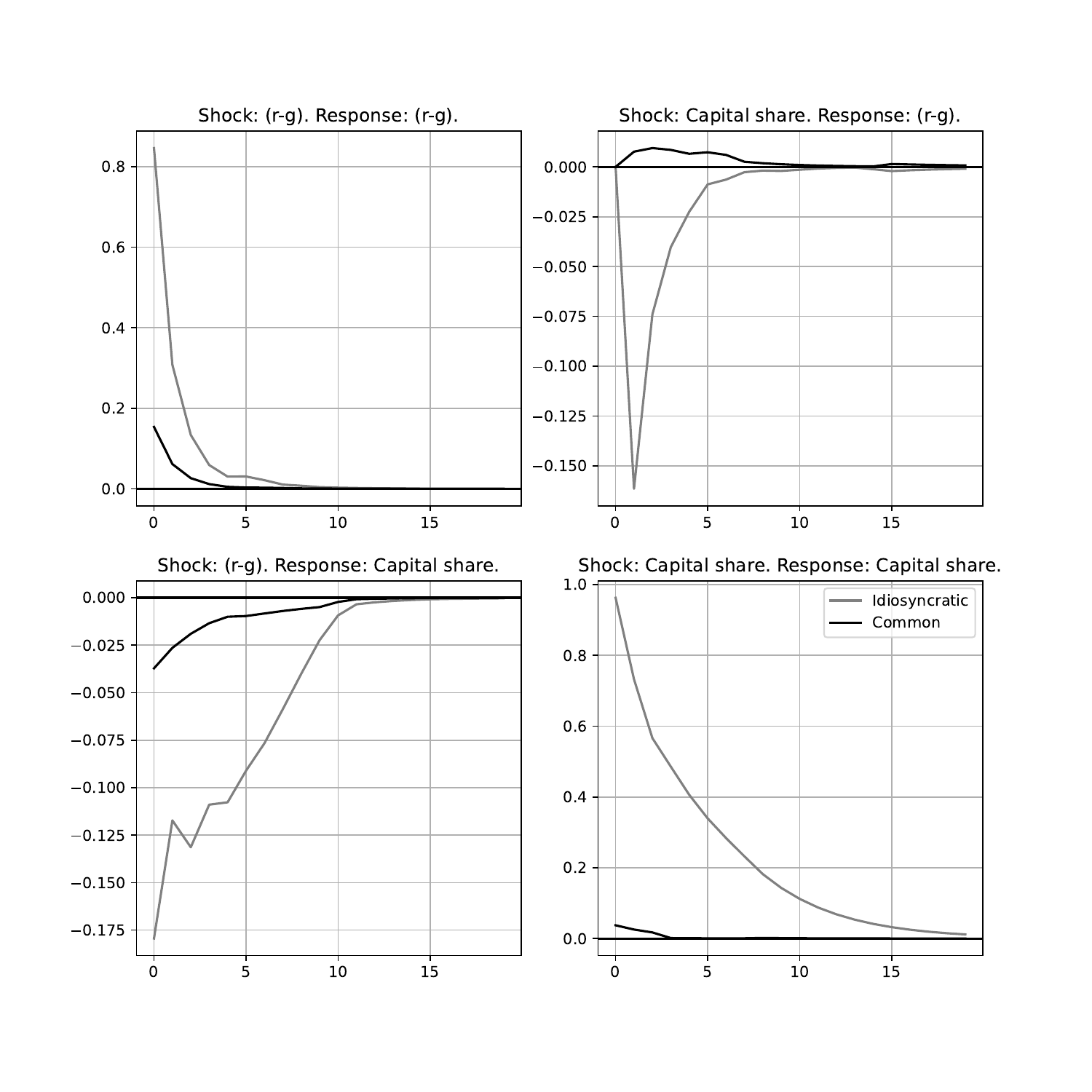}
\end{center}
\caption[Results of Model 2: Decomposition of median composite responses]{\textbf{Model 2: Decomposition of median composite responses.} Median responses can be decomposed into country-specific responses to common shocks and responses to idiosyncratic shocks through use of loading factors that denote the relative importance of common shocks for each country.}\label{fig:k-decomp}
\end{figure}

The median response of $r-g$ to capital-share shocks is negative (although not  statistically significant). This is intuitive since it is possibly reflecting diminishing marginal returns on capital\textemdash for any fixed growth rate, increased capital shares are likely to decrease returns and hence drive $r-g$ down on the new BGP. When compared to Model~1, idiosyncratic shocks are more dominant (see Figure~\ref{fig:k-decomp}).

\newpage

\newpage

\newpage

\begin{figure}[htp!] 
\begin{center}
    \includegraphics[width=0.5\textwidth]{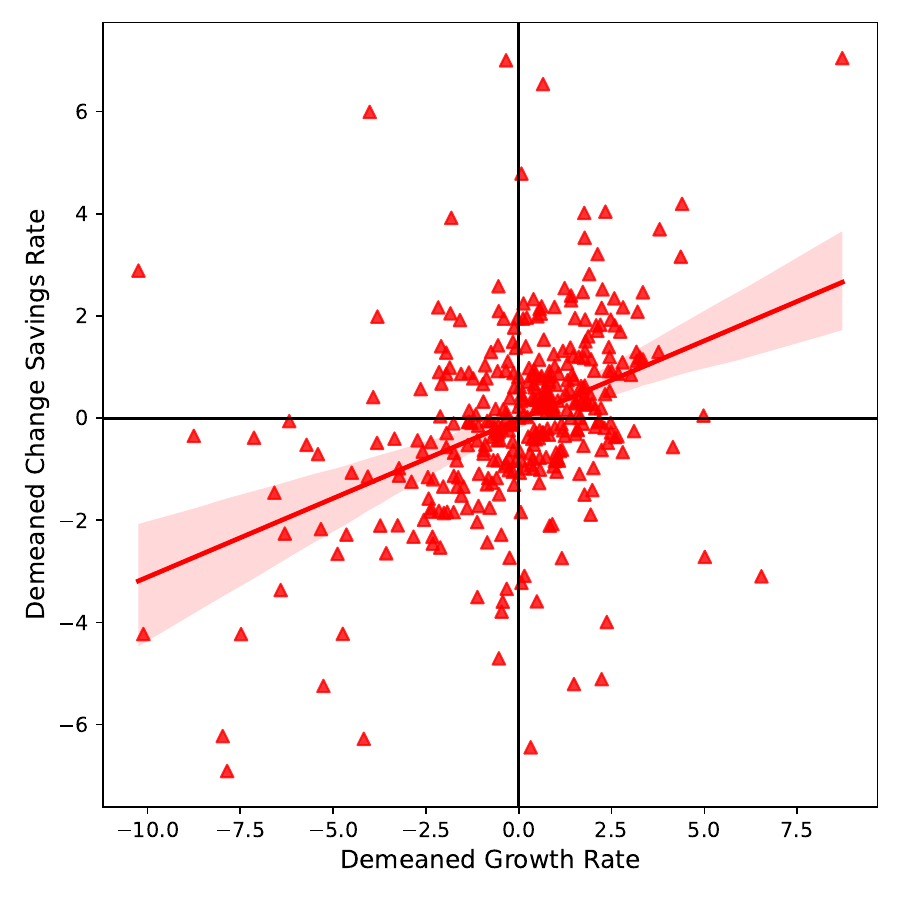}
\end{center}
\caption[Contemporaneous correlation between GDP growth and changes in the savings rate]{\textbf{Contemporaneous correlation between GDP growth and changes in the savings rate.} The sample refers to an unbalanced panel of 18 advanced economies ranging from 1981 to 2012. Variables are de-meaned to account for time-invariant, country-specific characteristics.}\label{fig:savings}
\end{figure}

This result seems unusual, as even in standard models one should expect the capital share to increase following $r-g$ shocks. However, the reason such evidence is absent is it is unreasonable to assume that the savings rate is exogenous to $r-g$. In fact, there are strong theoretical reasons why the savings rate should behave pro-cyclically. For instance, if a share of individuals move away from a borrowing constraint because of an expansion in the economy, the aggregate savings rate is expected to increase. In line with this theoretical rationale, changes in the savings rate in this sample are positively and statistically significantly correlated (Pearson's $\rho=0.40$) with GDP growth rates (see Figure~\ref{fig:savings}).

Since lower growth rates are associated with a decrease in savings rates, it is hard to know a priori the net effect of lower growth (which, all other things equal, are equivalent to a \textit{positive} $r-g$ shock) on the capital share. This happens because  if the decrease in the savings rate is large enough, it can offset any predicted effect of a $r-g$ shock on the capital share. To account for this, Model~3 includes the savings rate as an endogenous variable.

\begin{figure}[htp] 
\begin{center}
    \includegraphics[width=\textwidth]{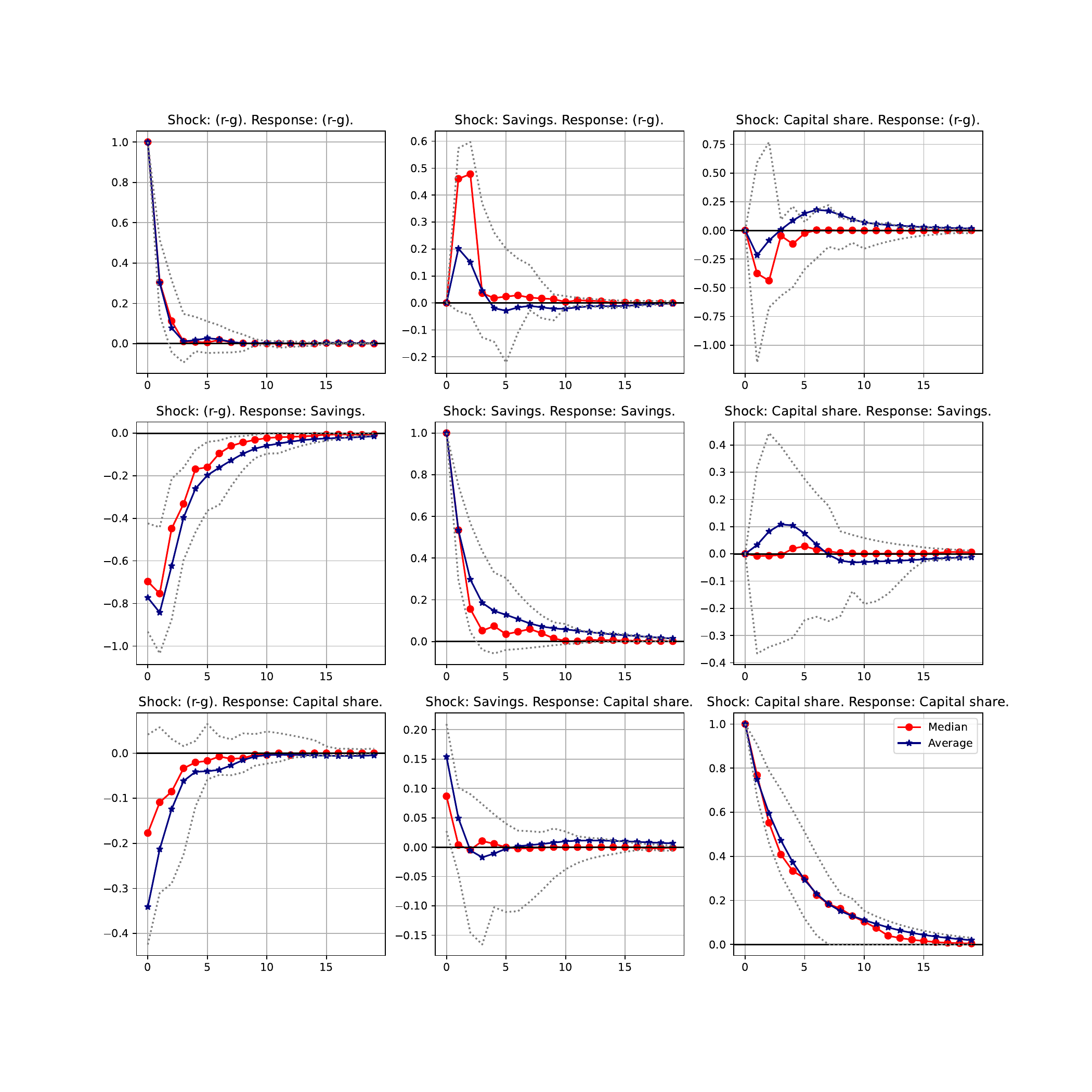}
\end{center}
\caption[Results of Model 3: Heterogeneous composite impulse responses across sample]{\textbf{Model 3: Heterogeneous composite impulse responses across sample.} The median and averages are shown as solid lines, interquartile ranges as dotted lines. All are calculated from the distribution of impulse response functions of the 18 cross-sections.}\label{fig:ksavings-dist}
\end{figure}

\begin{figure}[htp] 
\begin{center}
    \includegraphics[width=\textwidth]{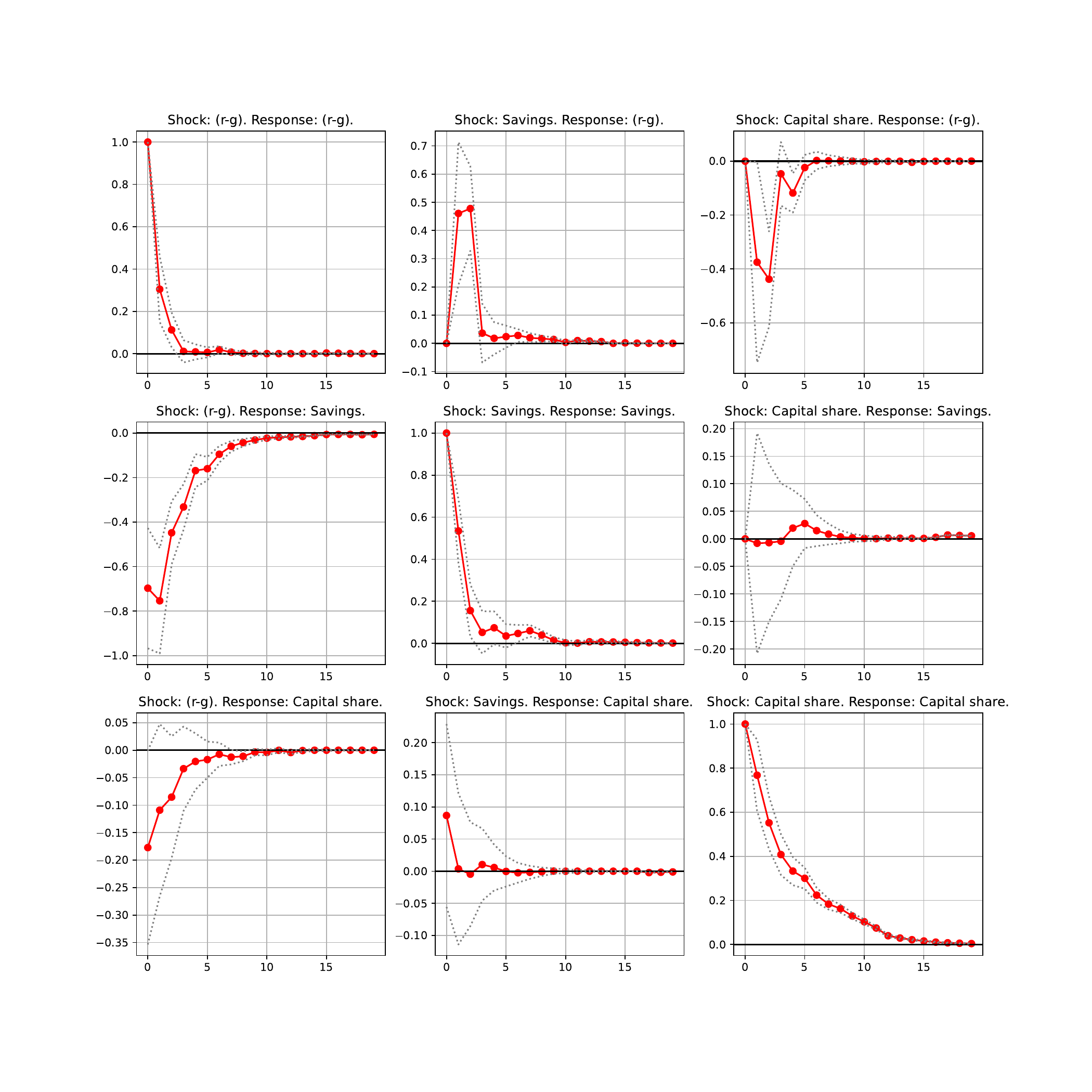}
\end{center}
\caption[Results of Model 3: Median composite responses and confidence intervals]{\textbf{Model 3: Median composite responses and confidence intervals.} Median response across a heterogeneous distribution of impulse response functions of 18 countries. Confidence intervals calculated from a resampling simulation with 1,000 repetitions. Since the distribution might change for each repetition in the simulation exercise, the confidence intervals do not represent the uncertainty around estimates for any particular country, but rather for the median estimate of the whole panel.}\label{fig:ksavings-ci}
\end{figure}

The relationship observed with basic contemporaneous correlations is also reflected in the impulse response functions. For at least 75\% of the countries, the savings rate responds negatively to a positive shock to $r-g$ (see Figure~\ref{fig:ksavings-dist}). The median response of the savings rate to $r-g$ shocks is large, negative, and statistically significant throughout the 20-year response horizon (see Figure~\ref{fig:ksavings-ci}). The responses of the savings rate help explain the counterintuitive result of the capital share responding negatively to such shocks.

\begin{figure}[htp] 
\begin{center}
    \includegraphics[width=\textwidth]{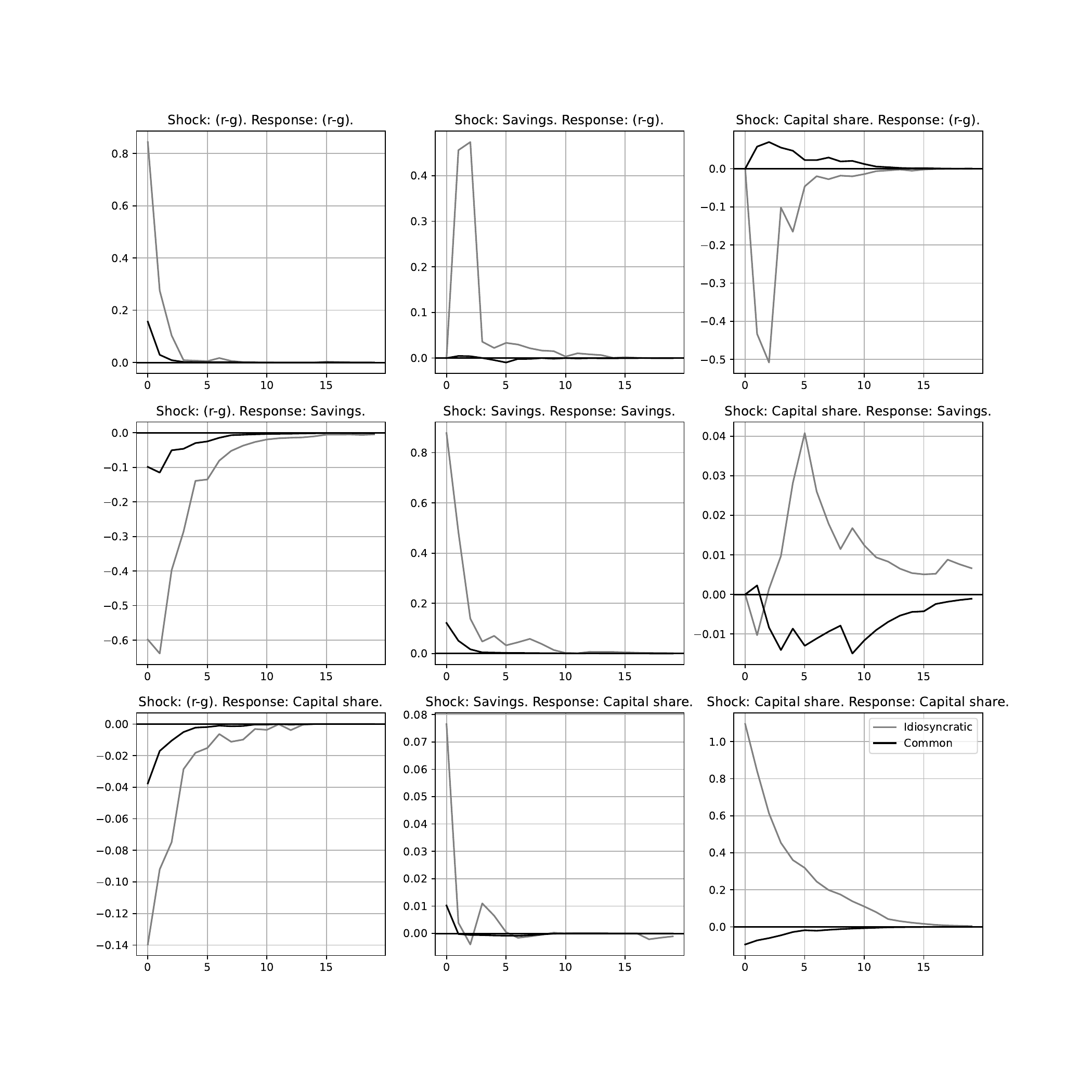}
\end{center}
\caption[Results of Model 3: Decomposition of median composite responses]{\textbf{Model 3: Decomposition of median composite responses.} Median responses can be decomposed into country-specific responses to common shocks and responses to idiosyncratic shocks through use of loading factors denoting the relative importance of common shocks for each country}\label{fig:ks-decomp}
\end{figure}

The median response of the capital share to $r-g$ shocks is not statistically significant, and as dynamics are incorporated, the median cumulative response of the capital share quickly nets out to somewhere close to zero (see Figure~\ref{fig:ksavings-ci}). Idiosyncratic shocks  contribute the most to capital-share dynamics in Model~3 (see Figure~\ref{fig:ks-decomp}). 

\section{Robustness}

\begin{figure}[htp] 
\begin{center}
    \includegraphics[width=\textwidth]{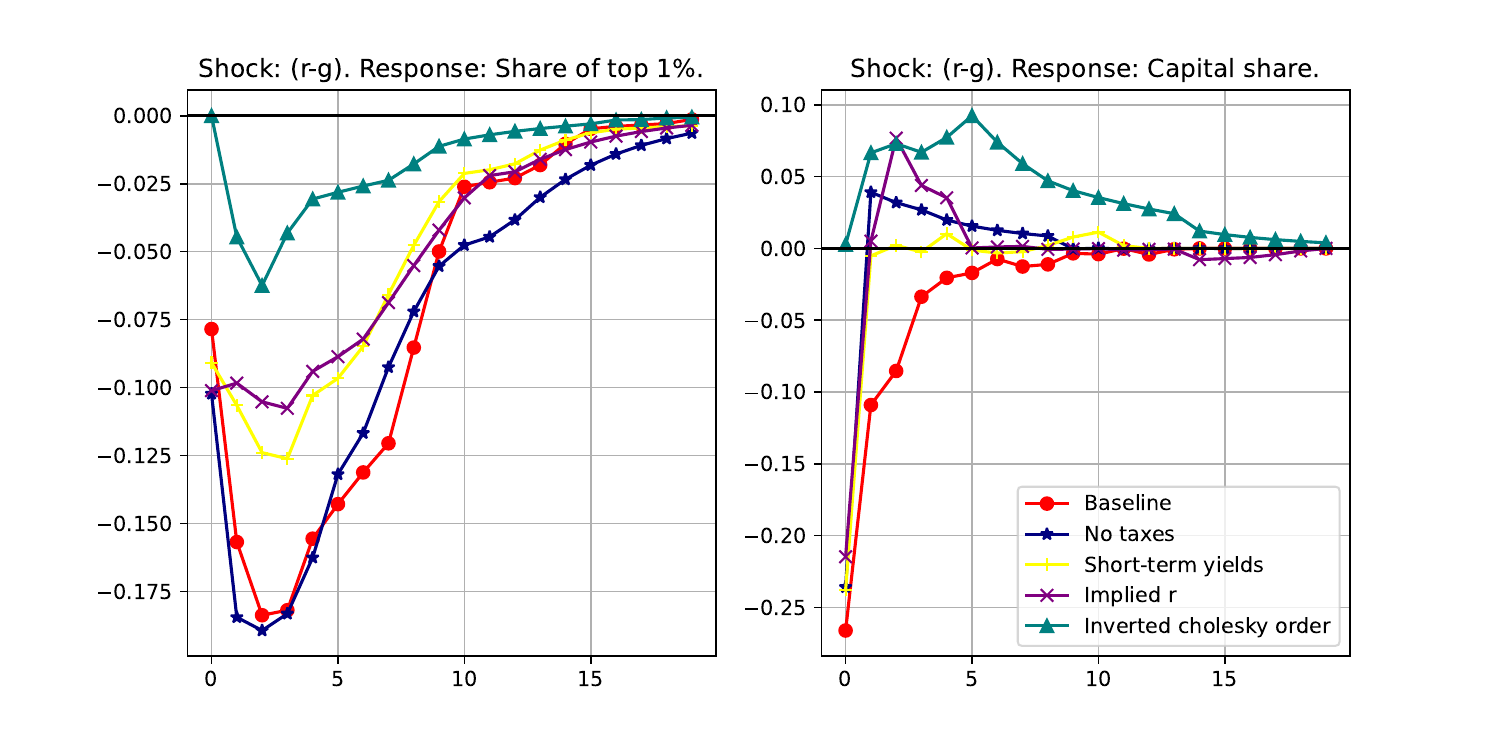}
\end{center}
\caption[Robustness checks]{\textbf{Robustness checks---re-estimation of Models~1 and 3: Median composite responses.} This figure compares median responses across a distribution of impulse response functions of 18 countries using different specifications.}\label{fig:robustness1}
\end{figure}

I run several alternative specifications to check the robustness of the results. I do that by modifying baseline Models~1 and 3. I present all alternative results in Figure~\ref{fig:robustness1}.

First, I run an alternative model ignoring tax rates. Tax rates are important for inequality, as pointed out by \citet{piketty5}, but it could be that deducting tax rates from returns is adding noise to the results. However, when re-estimating the models without taking taxes into account, the results are similar to the baseline. The median responses of the share of the top 1\% to $r-g$ shocks are still negative and in fact even more negative than in Model~1.

When replacing long-term bond yields with short-term yields (usually central bank policy rates), median responses are very similar to the baseline for the initial response horizons. In later years, responses net out to zero faster than in the baseline. 

I then re-estimate the models using an implied measure of $r$ from the national accounts. Deriving this measure is straightforward. Since capital share is the return on capital times the capital-to-income ratio ($\alpha = r K/Y$),   the return on capital is the capital share divided by the capital-to-income ratio ($r = \alpha Y/K$). After calculating implied returns on capital for all countries from the Penn World Tables, I deduct growth rates and re-estimate Model~1. Median responses still behave comparably to all other specifications, hovering close to zero.

Overall, the results are rather robust to different measures of $r$. There is no evidence that the share of the top 1\% responds positively to increases in the $r-g$ spread. Similarly, for the capital share, despite small differences, the results are virtually the same as in Model~3\textemdash with the savings rate playing a counterbalancing role after $r-g$ shocks and  with median capital-share responses going from negative to slightly positive before converging toward zero.
 
Finally, I run both models with an inverted Cholesky ordering---that is,  with the share of the top 1\% and the share of capital in national income, respectively, as those ordered first. It could be the case that the original restrictions imposed on the baseline specifications were hiding a positive effect of $r-g$ on the share of the top 1\% and the share of capital. However, in the alternative specification, such effects are very close to zero, which renders them both statistically and economically nonsignificant. Yet again, there is no evidence in support of Piketty's hypotheses of $r-g$ increasing the capital share and income inequality.
 
While the baseline conclusions are reasonably robust to all the different specifications estimated, there are several other possible extensions of the empirical models that could  provide additional robustness to the results. These include  using the Penn World Tables' estimates of time-varying depreciation rates, using measures of equity prices as a proxy for the return on capital, incorporating foreign savings, or using a different measure for inequality (such as  the Gini coefficient). 

\section{Discussion}
Piketty's conclusion that inequality will increase   rests on the  assumption that as growth decreases, driving the $r-g$ spread to increase, capital-to-income ratios will increase. However, the results from Models~2 and 3 fail to show robust positive responses of the capital share to shocks in $r-g$ and cast doubt on Piketty's conjecture. A possible reason  is that Piketty could be underestimating diminishing returns of capital\textemdash thereby overestimating the elasticity of substitution between capital and labor, whose empirical estimates tend to be much lower than what he assumes \citep{rognlie}. This relationship is illustrated in this paper by the negative median responses of $r-g$ to positive capital-share shocks. 

Another, less emphasized but equally important, problem with Piketty's conjecture is highlighted by \citet{krusell}, who argue that Piketty's predictions are grounded on a flawed theory of savings\textemdash namely, that the savings rate net of depreciation is constant\textemdash which exacerbates the expected increase in capital-to-income ratios as growth rates tend to zero. They present an alternative model in which agents maximize intertemporal utility, and they arrive upon a setting in which, on a BGP, the savings rate is pro-cyclical. By showing that the savings rate responds negatively to negative growth shocks (which, in turn, are translated into positive $r-g$ shocks) for at least 75\% of the countries in the sample, the results of Model~3 provide empirical support to Krusell and Smith's analysis.

\citet{piketty3} says in his online notes,  ``With $g = 0\%$, we’re back to Marx's apocalyptic conclusions,'' in which the capital share goes to 100\% and workers take home none of the output. While this is logically consistent with the model's assumptions, empirically there seem to be endogenous forces preventing that: Non-negligible diminishing returns on capital and pro-cyclical changes in the savings rate. These are two different ways in which the transmission mechanism from $r-g$ to the capital share might be blocked: With the former, at the limit the rate of return on capital tends to zero and there is no dynamic transmission; with the latter, if growth approaches zero, the savings rate  might ultimately become zero, offsetting any effect of lower growth on the capital share. They are, however, fundamentally different: The first concerns the production function and technological change, while the second has to do with the  behavior of capital owners over their life cycle.

Regarding inequality, the results from Model~1 contradict Piketty's prediction  that following exogenous shocks in $r-g$, inequality should increase. In fact, for at least 75\% of the countries in the sample, the result is disconfirmatory. These findings are in line with  results by \citet{acemoglu}, who find negative coefficients in single-equation panel models when regressing $r-g$ on the share of the top 1\%. This paper goes further, not only because the model takes all variables as endogenous, but also because it incorporates tax variability across countries. Additionally, by decomposing shocks into  common and idiosyncratic components, rather than using time dummies, Model~1 retains potentially important information about the effect of structural forces (for example, globalization) on these dynamics\textemdash which, as \citet{milanovic} argues, is a problem with Acemoglu and Robinson's analysis.

That a positive $r-g$ spread does not lead to higher inequality is not necessarily surprising. As illustrated by \citet{mankiw} with a standard model that incorporates taxation and depreciation, even if $r > g$, one can arrive at  steady-state inequality that does not spiral endlessly. \citet{milanovic2} explains that the transmission mechanism between $r > g$ and higher income inequality requires all of the following conditions to hold: (i) savings rates have to be sufficiently high; (ii) capital income needs to be more unequally distributed than labor income; and (iii) there needs to be a high correlation between drawing capital income and being at  the top of the income distribution. Through dynamic analysis, this paper shows that the negative responses of the savings rate to $r-g$ shocks violate the first condition, thereby preventing higher levels of inequality  compared to those observed before the increases in $r-g$.

These results suggest one should look to factors other than the spread between $r$ and $g$ to explain the increase in income inequality in advanced economies. And while Piketty's hypothesis has generated significant debate, recent empirical studies emphasize alternative drivers of inequality.

For example, \cite{mulas} highlight the increasing role of labor-income disparities rather than capital concentration. \cite{aghion2019innovation} link the increase in inequality with innovation and skill-biased technological change. If innovators are rewarded with higher incomes because of a temporary technological advantage (in  Schumpeterian fashion), inequality will be exacerbated. The authors show that innovation explains about a fifth of the higher inequality observed in the US since 1975.

\citet{imf}, after evaluating cross-country evidence, find that past changes in inequality in advanced economies are associated  most closely with two labor market changes: higher skill premia and lower union membership rates. \citet{imf2} also present results that correlate changes in labor market institutions, particularly lower union density, with increases in income inequality in advanced economies.
  
\citet{mare} and  \citet{greenwood} argue that changes in mating behavior helped exacerbate income inequality. The probability that someone will marry another person with a similar socio-educational background (called assortative mating) increased in tandem with the rise in income inequality in the US in   recent decades. The interaction between higher skill premia and more assortative mating has exacerbated  household income inequality because the gap between higher and lower earners became larger and couples became more segregated.

\citet{chong} use a dynamic panel to show that inequality tends to decrease as institutional quality improves. The underlying logic is that if the basic rules of economic behavior are not symmetrically enforced, the rich will have a higher chance to extract economic rents, thereby increasing inequality. \citet{acemoglu} make a similar argument. They say that economic institutions affect the distribution of skills in society, indirectly determining inequality patterns. However, \citet{bennett2017ambiguous} and \cite{bennett2024economic} review the economic literature and show that institutions, as proxied by economic freedom, have an ambiguous association with economic inequality and that this relationship is contingent on ``choice of country sample, time period and/or inequality measure used.''

\cite{chambers2022economic} argue that ``regulations and barriers to entry can exacerbate income inequality by generating compliance costs that disproportionately impact small businesses and low-income households,'' and they document that both income inequality and federal regulations have increased in the United States. Additionally, using a shift-share strategy, they show that states more exposed to federal regulations also experienced increases in income inequality. In the same spirit, \citet{haupt2023profits} shows that professions subject to occupational licensing exhibit a wage premium across all quantiles of the income distribution even after controlling for observable characteristics of individuals.

\citet{melo2022estimating} study the link between rent-seeking and income inequality. They show that income inequality (measured as share of the top 10\% or top 1\% in total income) tends to be higher in states and counties with a high proportion of workers in legal services and in local government (which they take as proxies for income inequality). Similarly, \citet{facchini2024growth} argue that the relationship between inequality and growth is mediated by rent-seeking institutions. Applying a two-way fixed-effects model  to a panel of countries, they find that the relationship is positive across countries with low-rent-seeking institutions and negative in countries with high-rent-seeking institutions. 

Another potential  explanation for  why it is  difficult to find  empirical evidence for the causal links set forth by Piketty is income-concentration mismeasurement. I rely on the preferred set of estimates by Piketty and his coauthors to test his hypothesis, those from the \textit{World Inequality Database}. However, an emergent literature shows that the measurement of top income can be highly uncertain. A large share of income flows is unobserved, and its estimation depends on assumptions regarding its distribution and valuation. For instance, \citet{auten2021top} present alternative series for top income shares, using different assumptions, and conclude that the after-tax income share for the top 1\% of incomes in the United States stayed flat in the last half century.  

None of the pieces of evidence provided in the papers listed above are definitive or mutually exclusive. Rather, they are most likely complementary explanations for the recent developments of  income inequality. What they  have in common is that none  point to capital-to-income ratios or $r-g$ dynamics as the driving cause  of inequality patterns.

Some years after the publication of \textit{Capital}, \citet{piketty4} himself recognized that the ``rise in labor income inequality in recent decades has evidently little to do with $r-g$, and it is clearly a very important historical development.'' He nonetheless emphasized that a higher $r-g$ spread will be important and will exacerbate future inequality changes.

However, the results in this paper show that this is likely not to be the case. The results corroborate the idea that recent inequality changes are not explained by $r-g$ and suggest that new shocks to $r-g$ will likely not lead to higher inequality, as there is no evidence that shocks to $r-g$ increase income inequality. Combined, the observed endogenous dynamics of $r-g$ and the share of the top 1\% and the capital share, respectively, cast doubt on the reasonableness of Piketty's prediction about inequality.

While this study provides robust evidence against Piketty's hypothesis, some limitations warrant discussion. First, the measurement of returns on capital is inherently challenging because of data limitations and assumptions about tax effects, asset heterogeneity, and depreciation. Future improvement in data measurement means that a re-evaluation of these empirical relationships could be warranted whenever new data become available.

Second, the time range of my sample  might be too short to capture the long-run relationship under scrutiny. While this is certainly possible, there are some limits to the criticism's bite. First, since the focus of the exercise is advanced countries, one expects them to be close to their respective BGPs. If so, it is more likely that the 30-year sample is long enough to capture the steady states and the effect of shocks that disturb such steady states. Second, \citet{acemoglu} run single-equation models using 10- and 20-year averages going back to the 19th century and find no statistically significant evidence that increases in $r-g$ increase the share of the top 1\% in  national income. Unfortunately, using those multi-decennial averages in an endogenous dynamic panel framework would render the identification restrictions so strong (namely, the restriction that one variable has no impact on another for 20 years) that it would make the entire exercise futile.

Last, unobserved variables, such as global supply chain shifts or technological adoption, might influence both $r-g$ and inequality dynamics, introducing potential omitted variable bias. While this paper improves upon previous works by adding heterogeneity across countries to the analysis and decomposing responses into common and idiosyncratic shocks, this  critique remains. One avenue for future research is trying to use more granular data within countries to better capture common shocks across arguably more similar units.

\section{Conclusions}

This paper is  Popperian: It tests interesting, logically consistent, and falsifiable hypotheses. It thereby contributes to the literature by empirically checking the  veracity of a very influential theory regarding income inequality patterns.

 I find no evidence to corroborate the idea that the $r-g$ gap drives the capital share in national income. There are endogenous forces overlooked by Piketty\textemdash particularly the cyclicality of the savings rate\textemdash that balance out predicted large increases in the capital share. On inequality, the evidence against Piketty's predictions is even stronger: For at least 75\% of the studied countries, the response of inequality to increases in $r-g$ has the opposite sign to that postulated by Piketty.

These results are robust to different calculations of $r-g$. Regardless of whether I measure  the real return on capital as long-term sovereign-bond yields, short-term interest rates, or implied returns from national accounting tables, the dynamics are the same. Excluding taxes does not alter the qualitative takeaways from the results either.

 The policy implications of high inequality can be important. There are different mechanisms, not evaluated in this paper, that suggest that inequality can be harmful to social welfare. For instance, \citet{alesina1994} develop an endogenous growth model in which capital taxes increase returns to labor, inducing a conflict between capital owners and workers regarding the optimal tax rate in the presence of inequality in capital ownership. \citet{banerjee1993} develop a model with capital market imperfections in which long-run prosperity depends on the initial wealth distribution. \citet{easterly2007} provides evidence that initial land endowment predicts future inequality and future inequality predicts development.  \citet{campante2007inefficient} draw up a theory and \citet{shughart2003rent} provide suggestive evidence that heterogeneity in the capabilities of interest groups can reinforce income inequality. Additionally, while most of the papers cited in the previous section are empirical, they too test alternative mechanisms and explanations for the observed increase in income inequality. 

Without knowing the underlying causes of the recent trends in increased income inequality, it is impossible to design policy actions to counter them. The solution advanced by Piketty and some of his coauthors is increased capital-income or wealth taxation \citep[see, for instance, the policy proposals in the book by][]{saez2019triumph}. However, if the links between his theory and inequality are found wanting, these policies might not lead to the desired outcomes.

Additionally, one should take into account the potential unintended consequences of these policies. Well-identified empirical studies show that workers may bear the burden of as much as 50\%--75\% of   total corporate tax incidence (\citeauthor{fuest2018higher}, \citeyear{fuest2018higher}; \citeauthor{malgouyres2023benefits}, \citeyear{malgouyres2023benefits}; \citeauthor{suarez2016benefits}, \citeyear{suarez2016benefits}). As the model developed by \citet{bethencourt2015political} shows, increased inequality might lead to a preference for higher tax rates as well as a deterioration of the tax base through tax evasion. Empirically, \citet{guyton2021tax} find that tax evasion is very large among top income earners in the US, with underreported income accounting for as much as 50\% of total income. Hence, even if one abstracts  from the question whether to model optimal policy through the lens of a ``benevolent social planner'' or a ``revenue-maximizing Leviathan'' \citep{brennan1980power}, investigating the implications of this causal link between $r-g$ and inequality is relevant for outlining optimal choices.

Knowing whether increases in $r-g$ lead to more inequality is very important, not only for economics as a science of human action but also for the policy implications. Inequality is a complex phenomenon, and its trends are   sluggish. It is certainly possible that the long-term relationships Piketty proposes exist and are simply not captured by the 30 years of data for the 18 advanced economies included in this sample. However, for a large range of countries, the evidence provided here suggests that if one is looking for  potential solutions to increasing income inequality, one should not focus on $r-g$, but elsewhere.

\bibliography{ref}

\newpage

\appendix
\numberwithin{equation}{section}
\section{Appendix: Data and Sources}
\label{appendix:appendix1}

\subsection{Countries Sample}

Samples are slightly different in Model~1 compared to Models 2 and 3 to account for dynamic (in)stability in different samples.

\textbf{Sample for Model 1}: Australia, Canada, Finland, France, Germany, Italy, Ireland, Japan, Korea, Netherlands, New Zealand, Norway, Portugal, Singapore, Spain, Sweden, Switzerland, the United Kingdom, and United States.

\textbf{Sample for Models 2 and 3}: Australia, Canada, Denmark, Finland, France, Italy, Japan, Korea, Netherlands, New Zealand, Norway, Portugal, Singapore, Spain, Sweden, Switzerland, the United Kingdom, and United States.

\subsection{Data Sources}

\begin{tabular}{lll}
\textbf{Variable} & \textbf{Scale} & \textbf{Source}\tabularnewline
\hline 
Share of the top 1\% & Percent of national income & World Top Income Database\tabularnewline
Capital share & Percent of national income & Penn World Tables \tabularnewline
Real GDP growth & Annual percent change & World Economic Outlook\tabularnewline
GDP deflators & Index & World Economic Outlook\tabularnewline
Gross national savings & Percent of national income & World Economic Outlook\tabularnewline
Corporate income tax rates & Percent & OECD's Tax Database\tabularnewline
10-YR sovereign bond yields & Percent & Bloomberg \& Haver Analytics\tabularnewline
Short-term interest rates & Percent & Haver Analytics \& IMF \tabularnewline
Implied return on capital & Percent & Derived from Penn World Tables \tabularnewline
U.S. 95th pctile wealth
 & Constant U.S. dollars & Congressional Budget Office \tabularnewline

\hline 
\end{tabular}

\newpage
\section{Appendix: Accumulated Impulse Response Functions}\label{appendix: ac_irfs}

\begin{figure}[htp!]
\begin{center}
        \includegraphics[width=\textwidth]{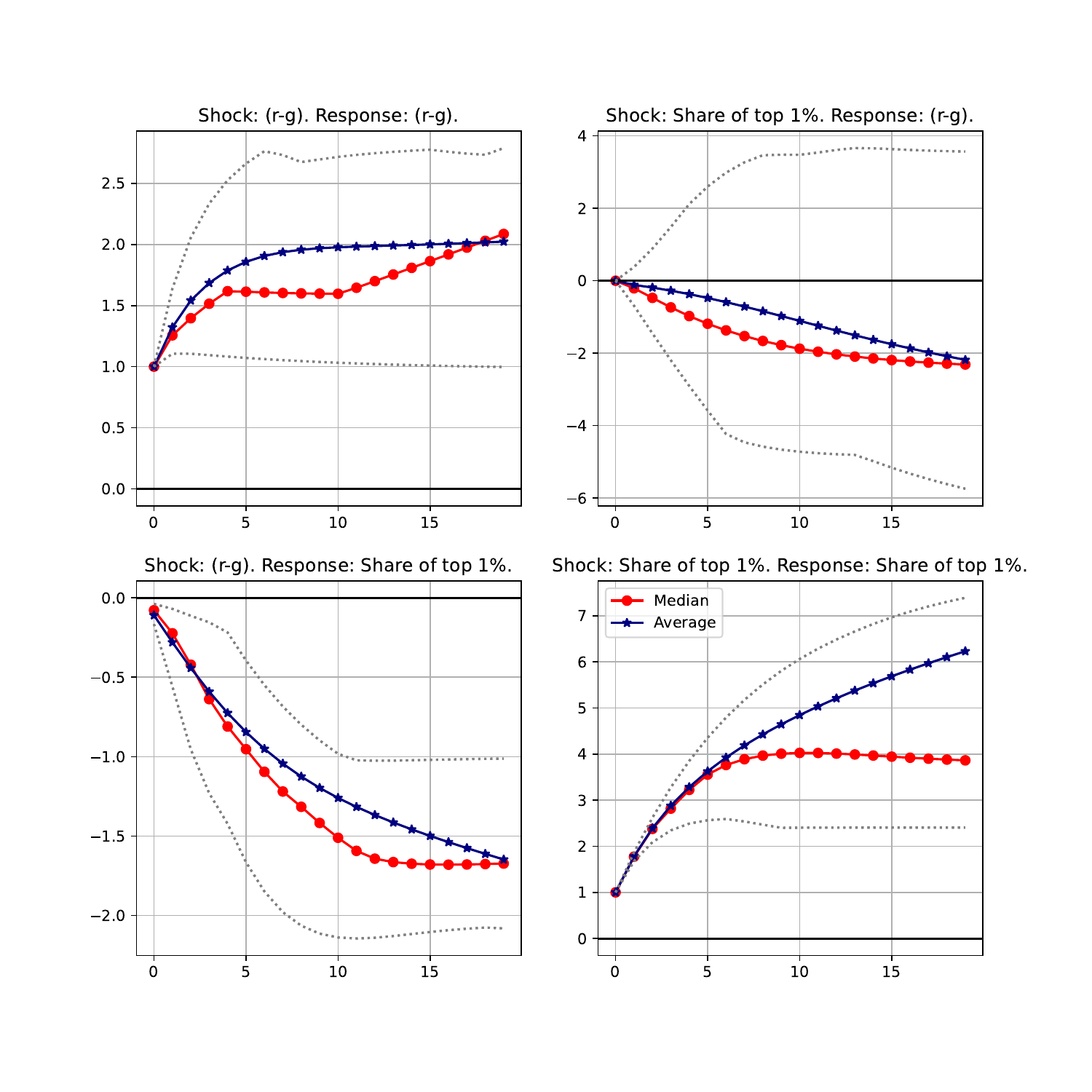}
\end{center}
\caption[Results of Model 1: Accumulated Heterogeneous composite impulse responses across sample]{\textbf{Model 1: Accumulated Heterogeneous composite impulse responses across sample}. The median and averages as solid lines. Interquartile ranges as dotted lines. All calculated from the distribution of impulse response functions of the 18 cross-sections.}\label{fig:top1-ac}
\end{figure}

\newpage

\begin{figure}[htp!]
\begin{center}
        \includegraphics[width=\textwidth]{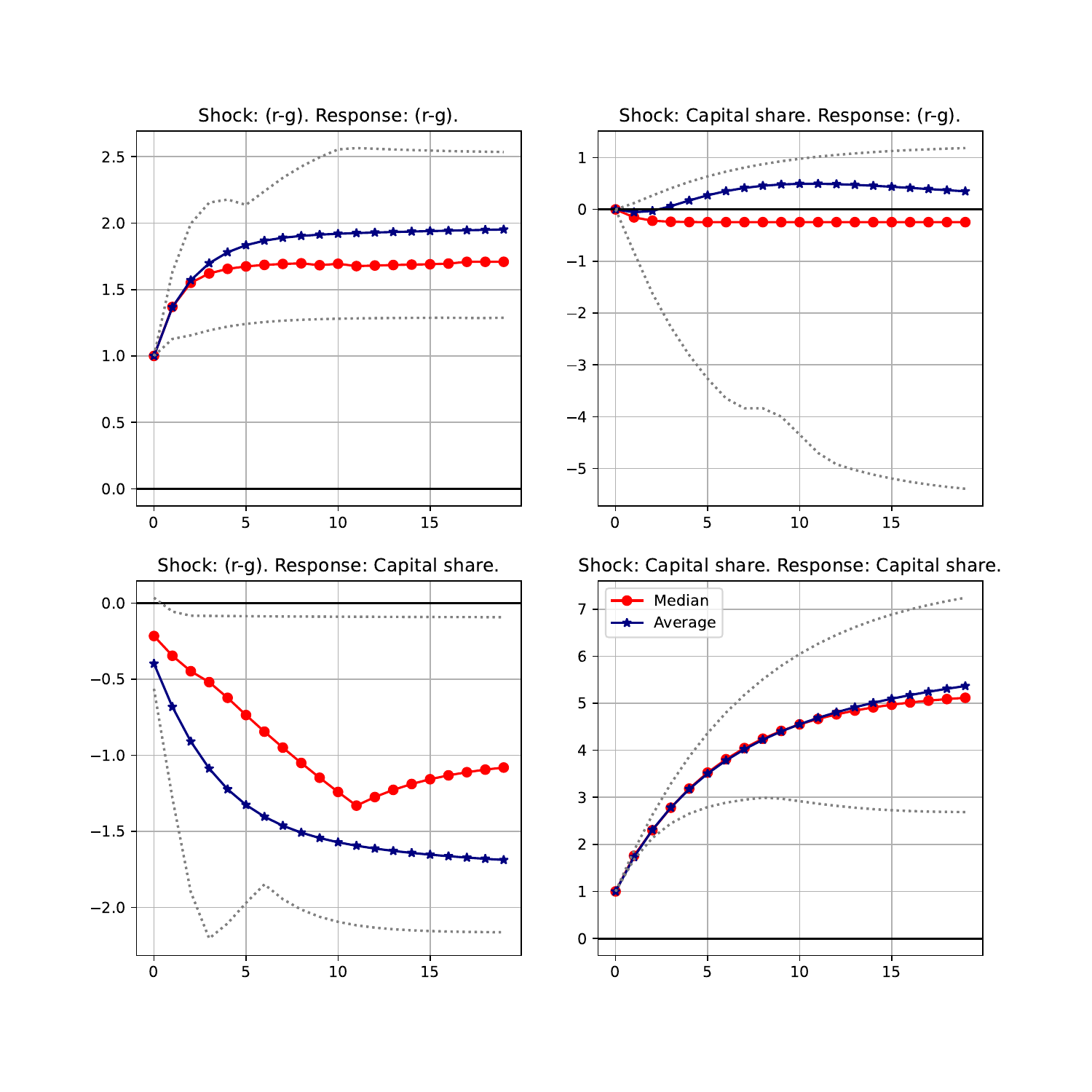}
\end{center}
\caption[Results of Model 2: Accumulated Heterogeneous composite impulse responses across sample]{\textbf{Model 2: Accumulated Heterogeneous composite impulse responses across sample}. The median and averages as solid lines. Interquartile ranges as dotted lines. All calculated from the distribution of impulse response functions of the 18 cross-sections.}\label{fig:k-ac}
\end{figure}

\newpage

\begin{figure}[htp!]
\begin{center}
        \includegraphics[width=\textwidth]{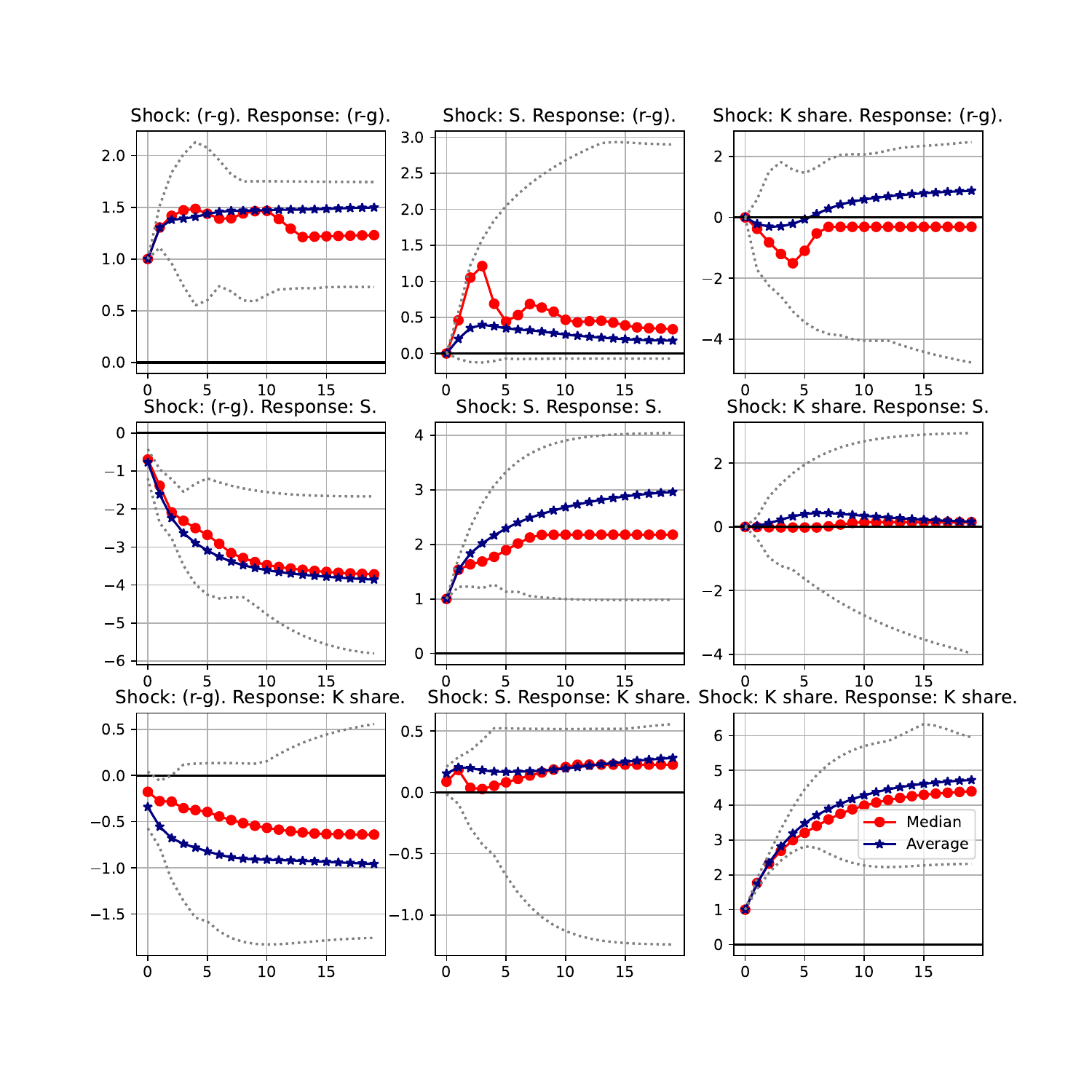}
\end{center}
\caption[Results of Model 3: Accumulated Heterogeneous composite impulse responses across sample]{\textbf{Model 3: Accumulated Heterogeneous composite impulse responses across sample}. The median and averages as solid lines. Interquartile ranges as dotted lines. All calculated from the distribution of impulse response functions of the 18 cross-sections.}\label{fig:ksavings-ac}
\end{figure}

\newpage
\section{Appendix: Empirical Methodology}\label{appendix: methodology}

For each member $i \in \{1, ..., N\}$ of an unbalanced panel, let $y_{i,t}$ be a vector of $M$ endogenous variables with country-specific time dimensions $t \in \{1, ..., T_i\}$. To control for individual fixed effects, I de-mean the data, resulting in $y^*_{i,t} = y_{i,t} - \bar{y}_{i} $, where $ \bar{y}_{i} \equiv T_i^{-1} \sum_{t=1}^{T_i} y_{i,t}    \forall   i$. The baseline model is: 
\begin{eqnarray}
    y^*_{i,t} &=& R_i(L)   y^*_{i,t-1} + u_{i,t} 
\end{eqnarray}

where $y^*_{i,t}$ is a $M$-dimensional vector of de-meaned stacked endogenous variables, $R_i(L)$ is a polynomial of lagged coefficients [$R_{i}(L) \equiv (\sum_{j=0}^{J_{i}} R_{j}^{i} L^{j})$] with country-specific lag-lengths $J_{i}$, $R^{i}_{j}$ is a matrix of coefficients, $u_{i,t}$ is a vector of stacked residuals, $J_i$ are selected based on appropriate criteria to assure that residuals approximate white noise. To allow for heterogeneous dynamics, I first estimate and identify reduced-form VARs for each country $i$:
\begin{eqnarray} \label{eq:vars}
   y^*_{1,t} &=& R_1(L)   y^*_{1,t-1} + u_{1,t} \nonumber \\ 
    &\vdots& \\
y^*_{N,t} &=& R_N(L)   y^*_{N,t-1} + u_{N,t} \nonumber
\end{eqnarray}

and then estimate another auxiliary VAR to recover \textit{common} dynamics. Common dynamics are captured by averages, across individuals, for each period ($\bar{y}^*_{t} \equiv N^{-1} \sum_{i=1}^{N} y^*_{i,t}$):
\begin{eqnarray} \label{eq:common}
    \bar{y}^*_{t} &=& \bar{R}(L)   \bar{y}^*_{t-1} + \bar{u}_{t}
\end{eqnarray} 

To transform reduced-form residuals in (\ref{eq:vars}) and (\ref{eq:common}) into their structural equivalents, I proceed in the following way. Note that structural residuals are related to reduced form residuals through a matrix of contemporaneous relationships $B_i$: $u_{i,t} = B_i e_{i,t} $ and $\bar{u}_{t} = \bar{B} \bar{e}_{t}$, respectively. First, I retrieve the variance–covariance matrices $\Sigma_{u_i} = E[ u_{i,t} u_{i,t}'], \Sigma_{\bar{u}} = E[ \bar{u}_{i,t} \bar{u}_{i,t}']$. Using the standard assumptions that structural residuals are uncorrelated ($e_{i,t} e_{i,t}' = I_m \quad \forall i$), I decompose the variance–covariance matrices into two triangular matrices $\Sigma_{u_i} = E[ B_i B_i'], \Sigma_{\bar{u}} = E[ \bar{B} \bar{B}']$.

Finally, I recover structural residuals as: $e_{i,t}  = B_i^{-1} u_{i,t}$, $\bar{B}^{-1}\bar{u}_{t} = \bar{e}_{t}$. 

With those at hand, the regression coefficient of composite on common shocks consistently estimate the loadings for the common versus idiosyncratic shocks. To do so, I run $mN$ OLS regressions to decompose the shocks into two terms:

\begin{eqnarray} \label{eq:loadings}
   e_{1,t} &=& \Lambda_1 \bar{e}_{t} + \tilde{e}_{1,t} \nonumber \\ 
    &\vdots& \\
   e_{N,t} &=& \Lambda_N \bar{e}_{t} + \tilde{e}_{N,t} \nonumber
\end{eqnarray}

where $e_{i,t}$ are \textit{composite shocks}, $\bar{e}_{i,t}$ are \textit{common shocks},  $\tilde{e}_{i,t}$ are \textit{idiosyncratic shocks}, and $\Lambda_i$ are $n$-by-$n$ diagonal matrices with country specific loadings (the coefficients from the OLS regressions) denoting the relative importance of common shocks for each country. Note that $\tilde{e}_{i,t}$ vectors are truly idiosyncratic, since they are by construction orthogonal to the shocks derived from the average dynamics shared by all panel members.

Subsequently, I use standard methods described in \citet{lutkepohl} to recover the matrices of composite responses to structural shocks [$A_i(L)$] for each country, which are shown below in the vector moving average representations of $N$ structural VARs:

\begin{eqnarray}
    y^*_{1,t} &=& A_1(L) u_{1,t} \nonumber \\ 
    &\vdots& \\
    y^*_{N,t} &=& A_N(L) u_{N,t} \nonumber
\end{eqnarray}

and then use the loading matrices estimated in (\ref{eq:loadings}) to decompose the composite responses into country-specific responses to common shocks and responses to idiosyncratic shocks:

\begin{eqnarray}
    A_1(L) &=& \Lambda_1 A_1(L) + (I - \Lambda_1 \Lambda_1') A_1(L)  \nonumber \\ 
    &\vdots& \\
    A_n(L) &=& \Lambda_n A_n(L) + (I - \Lambda_n \Lambda_n') A_N(L)  \nonumber
\end{eqnarray}

Equivalently, $A_i(L) = \bar{A}_i(L) + \tilde{A}_i(L)$, where $\bar{A}_i(L) \equiv \Lambda_i A_i(L)$ and $\tilde{A}_i(L) \equiv (I - \Lambda_i \Lambda_i') A_i(L)$. I then use the cross-sectional distribution of $A_i(L)$, $\bar{A}_i(L)$ and $\tilde{A}_i(L)$ to describe some properties of the collection of impulse response functions calculated, such as their medians, averages and  interquartile ranges.

After recovering the point estimates of all the impulse response functions, I calculate standard errors of medians through a re-sampling simulation repeating all the steps above 1000 times  (see Appendix \ref{appendix:appendix2} for details).

\newpage
\section{Appendix: Standard Error Simulation Algorithm}
\label{appendix:appendix2}
\numberwithin{equation}{section}

Before starting the simulation, I run the individual VARs in equation (\ref{eq:vars}) and collect predicted values ($\hat{y}_{i,t}$) and reduced form residuals ($u_{i,t}$). I then transform reduced-form residuals into their structural equivalents ($e_{i,t} = B_i^{-1}u_{i,t}$) and run the following algorithm:

\begin{enumerate}

    \item I follow these steps $k = 1000$ times:
    
    \begin{enumerate}

        \item I re-sample the structural residuals accounting for cross-sectional dependence. First, for each $i$, I use the loadings $\Lambda_i$ to recover idiosyncratic shocks $\{ \tilde{e}_{i,t}\}_{t=0}^T$, using the equations in  (\ref{eq:loadings}) to extract $\tilde{e}_{i,t} = e_{i,t} - \Lambda_i \bar{e}_t$, where  $\bar{e}_t$ denote the common shocks.
        
        \item Let $\dot{e}_{i,t}$ denote re-sampled structural residuals $e_{i,t}$. I resample common shocks $\dot{\bar{e}}_t$ and idiosyncratic shocks $\dot{\tilde{e}}_{i,t}$ and use these variables to construct resample structural residuals that incorporate the cross-sectional dependence structure of the panel as:

        \begin{equation}
            \dot{e}_{i,t} = \Lambda_i \dot{\bar{e}}_t + \dot{\tilde{e}}_{i,t}
        \end{equation}

        \item I then transform re-sampled structural residuals back into reduced-form residuals by pre-multiplying them with $B_i$: $\dot{u}_{i,t} = B_i \dot{e}_{i,t}$.
        
        \item I sum the predicted values of $y^*_{i,t}$ and re-sampled residuals to create pseudo-series for all members $N$. Let the pseudo-series be $\tilde{y}_{i,t} \equiv \hat{y}_{i,t} + \dot{u}_{i,t}.$
        
        \item I re-estimate the model with the pseudo-series: $\tilde{y}_{i,t} = \tilde{R}(L)   \tilde{y}_{i,t-1} + \tilde{u}_{i,t}$.
        
        \item I collect the matrices of responses, extract individual vectors for each pseudo-IRF, organize them into a matrix, and calculate the median response of this distribution.
        
        \item If this repetition $< k$, I go back to (a), otherwise, I move to step 2.
    
    \end{enumerate}
    
    \item After I repeat this procedure $k$ times the result will be a set of distribution matrices $D_m$ of simulated medians: 

\begin{equation}
    D_m = 
    \begin{bmatrix}
        \tilde{\rho}^1_{1m} & \hdots & \tilde{\rho}^k_{1m} \\
        \vdots & \ddots & \vdots \\
        \tilde{\rho}^1_{hm} & \hdots & \tilde{\rho}^k_{hm} 
    \end{bmatrix}_{hxk}
\end{equation}

where $m$ is the $m^{th}$ response variable, $h$ is the response horizon, and $k$ is the number of repetitions of the simulation exercise.

\item From $D_m$ I take the square root of the second moment of each row to build a vector of standard errors: 

\begin{equation}
    \sigma_{\rho_{m}} = 
    \begin{bmatrix}
        \text{Var}[\{\tilde{\rho}^1_{1m} , \hdots, \tilde{\rho}^k_{1m}\}]^{1/2} \\
        \vdots\\
        \text{Var}[\{\tilde{\rho}^1_{hm}, \hdots, \tilde{\rho}^k_{hm}\}]^{1/2} 
    \end{bmatrix}_{hx1}
\end{equation}
\end{enumerate}

\end{document}